\RequirePackage{rotating}
\documentclass[fleqn,usenatbib]{mnras}
\usepackage{graphicx}	
\usepackage{amsmath}	
\usepackage{amssymb}	
\usepackage{textcomp}
\usepackage{gensymb}
\usepackage{subfig}
\usepackage{hyperref}
\usepackage{threeparttable}
\usepackage{ulem}
\usepackage{hyperref}
\usepackage[version=4]{mhchem}
\usepackage{longtable}
\usepackage{rotating}
\usepackage{lscape}
\usepackage{newtxtext,newtxmath}
\usepackage[T1]{fontenc}
\usepackage{rotating}
\DeclareRobustCommand{\VAN}[3]{#2}
\let\VANthebibliography\thebibliography
\def\thebibliography{\DeclareRobustCommand{\VAN}[3]{##3}\VANthebibliography}

\title[Minute-cadence Observations of the TMTS V]{Minute-Cadence Observations of the LAMOST Fields with the TMTS: V. Machine Learning Classification of TMTS Catalogues of Periodic Variable Stars}

\author[Guo et al.]{
Fangzhou Guo,$^{1}$
Jie Lin,$^{2,3,1}$\thanks{E-mail: linjie2019@ustc.edu.cn}
Xiaofeng Wang,$^{1,4}$\thanks{E-mail: wang\_xf@mail.tsinghua.edu.cn}
Xiaodian Chen,$^{5}$
Tanda Li,$^{6,7,8}$
Liyang Chen,$^{1}$
Qiqi Xia,$^{1}$\newauthor
Jun Mo,$^{1}$
Gaobo Xi,$^{1}$
Jicheng Zhang,$^{9}$
Qichun Liu,$^{1}$
Xiaojun Jiang,$^{10,11}$
Shengyu Yan,$^{1}$
Haowei Peng,$^{1}$\newauthor
Jialian Liu,$^{1}$
Wenxiong Li,$^{12}$
Weili Lin,$^{1}$
Danfeng Xiang,$^{1}$
Xiaoran Ma$^{1}$
and Yongzhi Cai$^{13,14,15,1}$ 
\\
$^{1}$Department of Physics and Tsinghua Center for Astrophysics (THCA), Tsinghua University, Haidian District, Beijing 100084, China\\
$^{2}$CAS Key laboratory for Research in Galaxies and Cosmology, Department of Astronomy, University of Science and Technology of China, Hefei, 230026, China\\
$^{3}$School of Astronomy and Space Sciences, University of Science and Technology of China, Hefei, 230026, China\\
$^{4}$Beijing Planetarium, Beijing Academy of Sciences and Technology, Beijing 100044, China\\
$^{5}$CAS Key Laboratory of Optical Astronomy, National Astronomical Observatories, Chinese Academy of Sciences, Beijing 100101, China\\
$^{6}$Institute for Frontiers in Astronomy and Astrophysics, Beijing Normal University, Beijing 102206, China\\
$^{7}$Department of Astronomy, Beijing Normal University, Beijing, 100875, China\\
$^{8}$School of Physics and Astronomy, The University of Birmingham, UK, B15 2TT\\
$^{9}$Department of Astronomy, Beijing Normal University, Beijing, 100875, China\\
$^{10}$National Astronomical Observatories of China, Chinese Academy of Sciences, Beijing, 100012, China\\
$^{11}$School of Astronomy and Space Science, University of Chinese Academy of Sciences, Beijing, 100049, China\\
$^{12}$The School of Physics and Astronomy, Tel Aviv University, Tel Aviv 69978, Israel\\
$^{13}$Yunnan Observatories, Chinese Academy of Sciences, Kunming 650216, China\\
$^{14}$Key Laboratory for the Structure and Evolution of Celestial Objects, Chinese Academy of Sciences, Kunming 650216, China\\
$^{15}$International Centre of Supernovae, Yunnan Key Laboratory, Kunming 650216, China\\
}

\date{Accepted XXX. Received YYY; in original form ZZZ}

\pubyear{2024}

\begin{document}
\label{firstpage}
\pagerange{\pageref{firstpage}--\pageref{lastpage}}
\maketitle

\begin{abstract}
Periodic variables are always of great scientific interest in astrophysics. Thanks to the rapid advancement of modern large-scale time-domain surveys, the number of reported variable stars has experienced substantial growth for several decades, which significantly deepened our comprehension of stellar structure and binary evolution. The Tsinghua University–Ma Huateng Telescopes for Survey (TMTS) has started to monitor the LAMOST sky areas since 2020, with a cadence of 1 minute. During the period from 2020 to 2022, this survey has resulted in densely sampled light curves for \textasciitilde 30,000 variables of the maximum powers in the Lomb-Scargle periodogram above the 5$\sigma$ threshold. In this paper, we classified 11,638 variable stars into 6 main types using XGBoost and Random Forest classifiers with accuracies of 98.83\% and 98.73\%, respectively. Among them, 5301 (45.55\%) variables are newly discovered, primarily consisting of $\delta$ Scuti stars, demonstrating the capability of TMTS in searching for short-period variables. We cross-matched the catalogue with $Gaia$'s second Data Release (DR2) and LAMOST's seventh Data Release (DR7) to obtain important physical parameters of the variables. We identified 5504 $\delta$ Scuti stars (including 4876 typical $\delta$ Scuti stars and 628 high-amplitude $\delta$ Scuti stars), 5899 eclipsing binaries (including EA-, EB- and EW-type) and 226 candidates of RS Canum Venaticorum. Leveraging the metal abundance data provided by LAMOST and the Galactic latitude, we discovered 8 candidates of SX Phe stars within the class of "$\delta$ Scuti stars". Moreover, with the help of $Gaia$ color-magnitude diagram, we identified 9 ZZ ceti stars.
\end{abstract}





\section{Introduction}
\label{sec:intro}
\begin{figure*}
    \centering
    \includegraphics[width=1\textwidth]{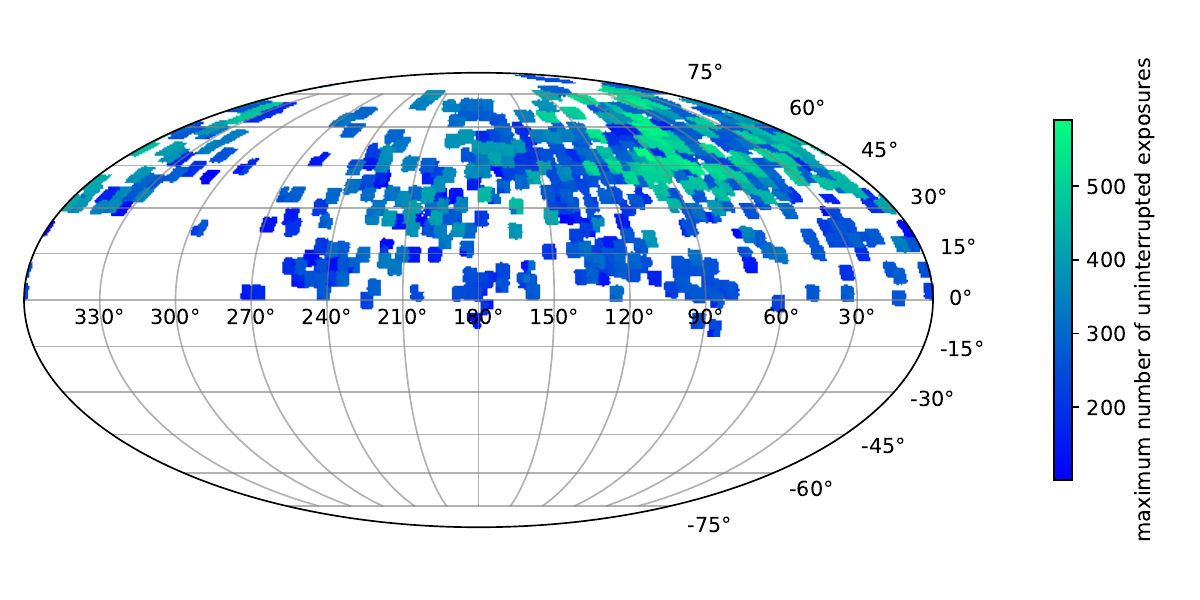}
    \caption{Sky map (in equatorial coordinates) of TMTS uninterrupted observations during 2020-2022. The color displays the maximum number of uninterrupted exposures within a single night, as in TMTS-II. We used the HEALPIX package (\url{http://healpix.sourceforge.net}) with NSIDE=128 to plot this sky map \citep{gorski2005healpix}.}
    \label{fig:skymap}
\end{figure*}
Since ancient times, people have been fascinated by changes occurring in the night sky. Large-scale surveys play a crucial role in the study of variables in the modern era. Over the past decade, missions like the All-Sky Automated Survey for Supernovae (ASAS-SN, \citealt{jayasinghe2019asas}), the Catalina Sky Survey (CSS, \citealt{drake2014catalina}), the Asteroid Terrestrial-impact Last Alert System (ATLAS, \citealt{heinze2018first}) among others offer unique opportunities to explore variable stars. The $Gaia$ DR2 provides a detailed catalogue of more than 1 million variable stars, including \textasciitilde 228 thousand RR Lyrae stars, \textasciitilde 151 thousand long-period variable stars (LPVs), \textasciitilde 147 thousand rotational variables, etc \citep{brown2018gaia,holl2018gaia}. $Gaia$ $G$, $G$$\mathrm{_{BP}}$ and $G$$\mathrm{_{RP}}$ integrated photometric band measurements as well as parallaxes with unprecedented accuracy substantially help to locate stars in the $Gaia$ color-magnitude diagram (CMD), which facilitates classification and further studies of stellar evolution history. The Optical Gravitational Lensing Experiment (OGLE) sky survey has started sky monitoring for more than 30 years. Initially designed as hunter for gravitational microlensing events and dark matter, the OGLE has obtained photometric datasets for millions of variables in the Galactic bulge, Galactic disk and the Magellanic Clouds \citep{udalski2015ogle}. In the foreseeable future, the Large Synoptic Survey Telescope (LSST, \citealt{ivezic2019lsst}) may discover billions of variable stars, potentially ushering in a transformative era in the field of astronomy. However, due to the long sampling cadences (e.g. $\geq$ 1 day), previous large-scale surveys are constrained in their ability to identify short-period variables. The advent of high-cadence surveys, such as the ZTF high-cadence Galactic Plane Survey with a cadence of 40 sec \citep{kupfer2021year}, has brought groundbreaking opportunities to the identification of variables with periods below several hours, such as ultra-compact binaries (UCBs) and compact pulsators. 

Study of variable stars is an essential part in time-domain astronomy, which focuses on transients and violent outbursts in the universe. Pulsating stars are a vital category among variable stars. Situated within the instability strip, pulsating stars mainly consist of Cepheids, $\delta$ Scuti stars, RR Lyrae stars, ZZ Ceti stars and LPVs. Mainly triggered by the $\kappa$ mechanism, these pulsators follow period-luminosity ($P-L$) relations, among which the well-defined $P-L$ relations of Cepheids \citep{leavitt1912periods} rendering them invaluable for measuring cosmic distances and aiding in the determination of the Hubble constant \citep{sandage2006hubble,riess2018milky}. In the big family of pulsating variables, stars manifest significant discrepancies in their periods. $\delta$ Scuti stars are a subclass of pulsating stars which undergo multi-mode pulsations \citep{breger2000delta} with periods below 0.3 d. This bound is somewhat arbitrarily set to distinct them from classical pulsators such as RR Lyrae stars and Cepheids. Pre-main-sequence A-type stars are the fastest known $\delta$ Scuti pulsators \citep{holdsworth2014high}, among which HD 34282 has the shortest period ($P$ = 18.12 min) ever observed \citep{amado2004pre}. The study of $\delta$ Scuti stars would answer some long-standing questions in stellar physics, such as the pulsator fraction within the $\delta$ Scuti instability strip as well as the determination of the edges of the instability strip \citep{murphy2019gaia}. On the other hand, the periods of Cepheids are usually a few days. For LPVs like mira variables which are characterized by very red colors, periods can extend to hundreds or even thousands of days \citep{iwanek2022ogle}. This vast disparity in periods reflects the abundance and diverse nature of pulsating stars.

Eclipsing binaries are another important subclass of transients discovered by wide-field survey projects. Binaries play a crucial role in understanding stellar evolution history (e.g. common envelope evolution, mass exchange and stellar winds, \citealt{taam2000common,plavec1968mass,theuns1993wind}). In addition, they could be progenitor of Type Ia Supernovae \citep{rebassa2019double}, which is of significant importance in cosmology. Periods of binaries range from several hours to a few days, while short-period binaries have been getting increasing attention recently, since the mergings of UCBs, whose period are typically less than an hour \citep{chen2020detectability}, are the most prominent gravitational wave source \citep{postnov2014evolution}. Observed by the LIGO-Virgo detector network, the gravitation-wave signal GW170817 is explained as a binary neutron star merger, while GW190814 is generated by a compact binary coalescence involving a black hole and a compact object \citep{abbott2017gw170817,abbott2020gw190814}. Several short-period eclipsing binaries, such as the AM CVn stars SDSS J0926+3624 ($P$ = 28 min, \citealt{copperwheat2011sdss}), the double-white-dwarf binary SDSS J0651+2844 ($P$ = 12.75 min, \citealt{brown201112}) and ZTF J1539+5027 ($P$ = 6.91 min, \citealt{burdge2019general}), have periods considerably shorter than the 0.22 day short-period limit of contact eclipsing binaries \citep{rucinski1992can}, posing a new challenge to this short-period limit.

In the past decade, machine learning technique becomes increasingly prevalent in astronomy, with the astronomical data underwent exponential expansion, (e.g., \citealt{richards2011machine,bloom2012automating,kim2016package,naul2018recurrent,hosenie2020imbalance}). Holding great potential to enhance our interpretation of vast and complex data, machine learning algorithms effectively resolve regression, classification, clustering and dimension reduction problems. In practice, machine learning techniques show particular promising in tackling intriguing problems such as galaxy identification \citep{krakowski2016machine}, variable star classification \citep{jayasinghe2019asas,chen2020zwicky} and exoplanet exploration \citep{nixon2020assessment}. 

The Tsinghua University–Ma Huateng Telescopes for Survey (TMTS), consisting of four optical telescopes, is located at the Xinglong Station of NAOC. TMTS has a field of view (FoV) up to 18 deg$\mathrm{^2}$ \citep{zhang2020tsinghua}, covering a wavelengths range from 400 to 900 mm. For an exposure time of 60s, TMTS can reach \textasciitilde 19.4 mag in white light (or Luminous filter) at 3$\sigma$ detection limit. To ensure the quality of our analysis, we only picked the light curves (LCs) with at least 100 uninterrupted measurements (\citealt{lin2022minute}, hereafter TMTS-I). TMTS has already yielded progress in several aspects, including the identification of over 1000 short-period variables (i.e., $P$ < 2 hr, \citealt{lin2023minute}, hereafter TMTS-II), the record of 125 flare stars \citep{liu2023minute}, the discoveries of a 18.9-min blue large-amplitude pulsators (BLAPs, \citealt{lin202318}) and a 20.5-min sdB binary \citep{lin2023sevenearthradius}. In this paper, we search for periodic variables ($P$ < 7.5 hr) using two advanced algorithms (XGBoost and Random Forest) in the TMTS Catalog of Periodic Variable Stars.
Observations and data are introduced in Section \ref{sec:data}, and the method and result of the classification are discussed in Section \ref{sec:class}. Section \ref{sec:cata} explains the characteristics of the catalogues. We discuss the overall catalogue and each type of variables in detail in Section \ref{sec:discussions}. A summary of our work is given in Section \ref{sec:summary}. 


\section{Data}
\label{sec:data}

Since 2020, TMTS has monitored 449 LAMOST/TMTS plates by the end of 2022, covering a total of 6977 deg$\mathrm{^2}$ of the sky, which are shown in Figure \ref{fig:skymap}. This effort leads to the production of 19,099,266 uninterrupted LCs with more than 100 epochs. We adopted the Lomb-Scargle periodogram (LSP; \citealt{lomb1976least,scargle1982studies}) to analyze these LCs. Using the distribution of modified false-alarm probability (FAP) as adopted in papers of TMTS-I and -II, we calculated the maximum powers in LSP (hereafter $LSP\mathrm{_{max}}$) and determined the 5$\sigma$ and 10$\sigma$ thresholds for each source. Among the millions of sources detected, the numbers of those with $LSP\mathrm{_{max}}$ $>$ 5$\sigma$ and $>$ 10$\sigma$ thresholds are $>$ 30000 and $>$ 8000, respectively. The TMTS LCs are cross-matched with $Gaia$ DR2 (the pipeline system of TMTS cross-matched DR2) and LAMOST DR7 to obtain some crucial photometric and spectroscopic parameters, such as dereddening color $(B\mathrm{_p}-R\mathrm{_p})_0$, effective temperature $T$$\mathrm{_{eff}}$ and spectral type. We derived the G-band absolute magnitude ($G$$\mathrm{_{abs}}$) of variables with reliable parallax (i.e., $\varpi/\sigma_\varpi \geq 5.0$) as in paper TMTS-II. 

To ensure the accuracy of our classifications of the TMTS periodic variables, we first cross-matched the TMTS LCs with the variable sources of the International Variable Star Index (VSX, \citealt{watson2006international})\footnote{The current version was updated on May 2 2023.}, which is a database providing comprehensive information of over 2200,000 known and suspected variable stars. This cross match indicates that 6801 sources have been recorded and identified by the VSX. To augment the dataset, a manual examination was conducted, resulting in another 1015 sources with assigned classifications. Among the ($6801+1015=$) 7816 identified sources, 6146 have $LSP\mathrm{_{max}}$ above the 10$\sigma$ threshold. 

We visually inspected all the LCs of the 6146 sources and excluded some low-quality (low SNR ratio, i.e. SNR $<$ 20) LCs. It is worth noting that classifying a variable star based solely on the information extracted from the LCs can be challenging. Incorporating supplementary information, especially color and absolute magnitude, can significantly enhance the reliability of classification. The $Gaia$ color-magnitude diagram (CMD), a close counterpart to the Hertzsprung–Russell diagram, is a useful tool for the classification, as it enables the segregation of variables into distinct regions based on their characteristics \citep{eyer2019gaia}. For instance, $\delta$ Scuti stars locate in the instability strips, while eclipsing binaries can appear anywhere in the CMD. 

Considering the importance of $Gaia$ $(B\mathrm{_p}-R\mathrm{_p})_0$ and $G$$\mathrm{_{abs}}$, we exclusively deemed the classification of sources with these information to be reliable. A total of 4506 sources were finally selected as the labelled dataset (hereafter Dataset-I), which serves as the training set and test set to train the classifiers and evaluate their performances. Dataset-I contains 1120 typical $\delta$ Scuti stars (DSCT), 256 high-amplitude $\delta$ Scuti stars (HADS), 49 RS Canum Venaticorum (RS), and 3081 eclipsing binaries. The eclisping bianries includes 2985 W UMa type (EW), 42 Algol type (EA) and 54 $\beta$ Lyr type (EB). ``DSCT'' refers to typical $\delta$ Scuti stars which include subtypes of DSCT, DSCTc (obsolete VSX type designation, low-amplitude $\delta$ Scuti stars with V-band light amplitude $<$ 0.1 mag) and DSCTr (VSX type designation, a subtype of $\delta$ Scuti stars in ASAS-3). SX Phe variables are metal-poor pulsating sub-dwarfs which resemble $\delta$ Scuti stars phenomenologically. As SX Phe variables can hardly be identified through LCs, they are included in the class ``DSCT'' as well \citep{soszynski2021over}. We will discuss candidates of SX Phe variables using the metal abundance data provided by LAMOST and the Galactic latitude in Section \ref{sec:cata}. We use ``HADS'' to represent HADS and HADS(B) (VSX type designation, first/second overtone double-mode HADS). The criteria that distinguish HADS from DSCT is that the former light-variation amplitude exceeds 0.3 mag in the $V$ band. While this threshold may appear somewhat discretionary, HADS and DSCT exhibit divergent characteristics in light of the LC shapes, rotation speed, evolutionary stages and so on (\citealt{chang2013statistical,chen2020zwicky,mcnamara2011delta}). We therefore regard them as two distinct groups. To avoid ambiguity, we use ``$\delta$ Scuti stars'' to encompass the entire category (including typical $\delta$ Scuti stars and high-amplitude $\delta$ Scuti stars), while ``DSCT'' is used to denote typical $\delta$ Scuti stars and ``HADS'' specifically designates high-amplitude $\delta$ Scuti stars. The sample sizes for other types are insufficient (i.e., number $<$ 30) to establish datasets with adequate statistical robustness without risking overfitting, even with carefully-designed over-sampling algorithms. This overfitting could arise from that the models misinterprete the specific characteristics of individual light curves as general traits of the entire class. In the future, we intend to employ novelty detection algorithms for the identification of such variables.

We randomly checked 300 unlabeled LCs, finding a relatively low false-positive rate in the 10$\sigma$ threshold LCs. However, over 70\% of the 5$\sigma$ threshold LCs were recognized to be non-astrophysically variable sources, as a typical concern in high-cadence surveys \citep{kupfer2021year}. We thus retained all the 10$\sigma$ threshold LCs, while applying an additional minimum threshold of $LSP\mathrm{_{max}}$ = 30 for the 5$\sigma$ threshold LCs. This selection process yields \textasciitilde 8000 unlabeled LCs, referred to as Dataset-II. 

Dataset-I and II only include limited kinds of variables due to the observation strategy of TMTS (see paper TMTS-I). The upper period bound of TMTS is \textasciitilde 7.5 h, preventing us from discovering variables with longer periods, such as Cepheids and LPVs. In fact, the longest period detected in Dataset-I and -II is 6.21 h. Periods of Cepheids are typically above 1 d, far exceeding the upper period limit of TMTS. RR Lyrae stars are another frequently observed stars with periods of 0.2–1.0 d. A considerable portion of RRc variables having periods below 0.3 d (7.2 h), falls within the detection range of TMTS. However, they are old and commonly present in globular clusters located around the Galactic core \citep{soszynski2014over}, which is not covered by TMTS, resulting in their scarcity in our catalogue. 

\section{Classification}
\label{sec:class}

\subsection{Method}

\begin{table*}
	\centering
	\caption{Features of the XGBoost and RF classifier.}
	\label{tab:features}
	\begin{tabular}{cp{11cm}c} 
		\hline
            \hline
		Feature & Description & Reference\\
		\hline
            Period & Photometric period determined through the Lomb–Scargle periodogram. & \\
            $(B\mathrm{_P}-R\mathrm{_P})_0$ & $(G$$\mathrm{_{BP}}$-$G$$\mathrm{_{RP}})_0$, dereddened color provided by $Gaia$ DR2. & \\ 
            $M\mathrm{_G}$ & G-band (330 nm to 1050 nm) absolute magnitude provided by $Gaia$ DR2. & \\
            Amplitude & Peak-to-peak amplitude obtained from the fourth-order Fourier fitting of the LCs. & \\  
            Cusum & The range of the cumulative sum of the LCs. Cusums of LCs with longer-term variability are usually larger. & \cite{kim2014epoch}\\ 
            Eta & Measures the degree of trends in a long-term baseline, which is useful in separate variables with different periods. & \cite{kim2014epoch}\\
            Hl\_amp\_ratio & Ratio between higher and lower amplitude than average. Hl\_amp\_ratios for EAs are higher by definition. & \cite{kim2016package}\\
            Kurtosis &  The fourth standardized moment of the distribution, measuring the tailedness of the distribution. Kurtosis=$\frac{\frac{1}{n}\sum_{i=1}^{n}(x_i-\bar{x})^4}{(\frac{1}{n}\sum_{i=1}^{n}(x_i-\bar{x})^2)^2} - 3 $ & \\
            Quartile$_{31}$ & Difference between 75\% percentile and 25\% percentile of the LCs. & \cite{kim2016package}\\
            Shapiro\_w & Shapiro-Wilk test for normalization. & \cite{kim2016package}\\
            Skewness & The third standardized moment of the distribution, measuring its asymmetry. Skewness=$\frac{\frac{1}{n}\sum_{i=1}^{n}(x_i-\bar{x})^3}{(\frac{1}{n}\sum_{i=1}^{n}(x_i-\bar{x})^2)^3/2}$& \\
            Stetson\_k & Stetson K index, describing the shape of the LCs. & \cite{kim2014epoch}\\
            Weighted\_mean & Weighted mean of the LCs, the weight of each data point in the LCs is inversely proportional to its measurement error, which assigns higher weights to data points with smaller errors. & \\
            Weighted\_std & Weighted standard deviation of the LCs. & \\
            Phase\_eta & Eta for phase-folded LCs. & \cite{kim2014epoch}\\
            Phase\_cusum & Cusum for phase-folded LCs. & \cite{kim2014epoch}\\
            Slope\_per10 & 10\% percentile of slopes of a phase-folded LC. & \cite{long2012optimizing}\\
            Slope\_per90 & 90\% percentile of slopes of a phase-folded LC. & \cite{long2012optimizing}\\
            R$_{21}$ & $a_2$ / $a_1$, 2nd to 1st amplitude ratio obtained from the fourth-order Fourier fitting.\\
            R$_{31}$ & $a_3$ / $a_1$, 3nd to 1st amplitude ratio obtained from the fourth-order Fourier fitting.\\
            R$_{41}$ & $a_4$ / $a_1$, 4nd to 1st amplitude ratio obtained from the fourth-order Fourier fitting.\\
            $\Phi$$_{21}$ & $\Phi_2$ - 2$\Phi_1$, the phase difference between 2nd and 1st phase obtained from the fourth-order Fourier fitting.\\
            $\Phi$$_{31}$ & $\Phi_2$ - 3$\Phi_1$, the phase difference between 3rd and 1st phase obtained from the fourth-order Fourier fitting.\\
		\hline
	\end{tabular}
\end{table*}

\begin{figure}
    \centering
    \includegraphics[width=0.5\textwidth]{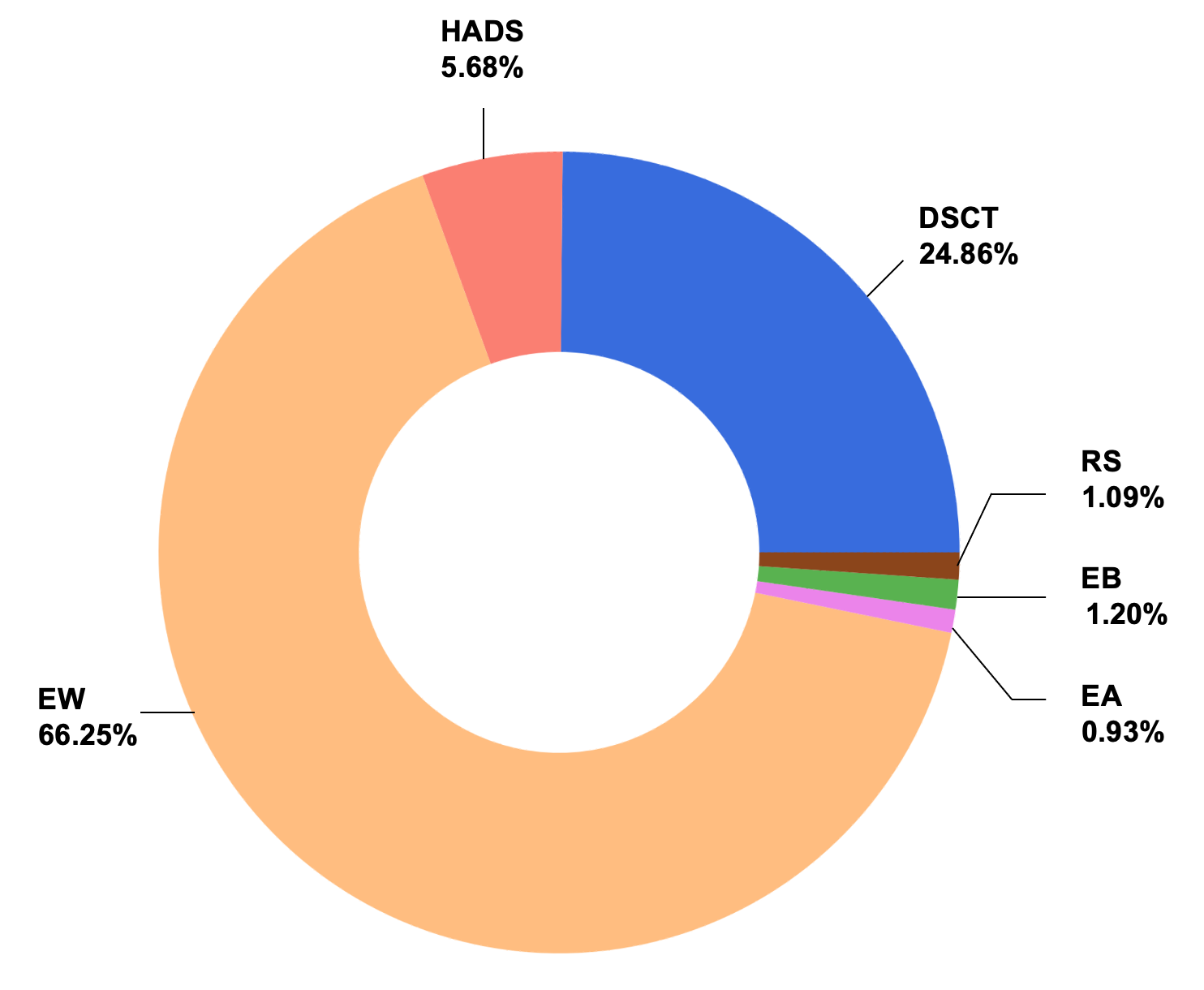}
    \caption{Pie chart of the distribution of variable stars within Dataset-I, which is heavily imbalanced. EWs account for 66.25\%, while EAs comprise a mere 0.93\%, underscores the pronounced disparity of over 70-fold between these two groups.}
    \label{fig:pie}
\end{figure}

\begin{figure*}
    \centering
    \includegraphics[width=1\textwidth]{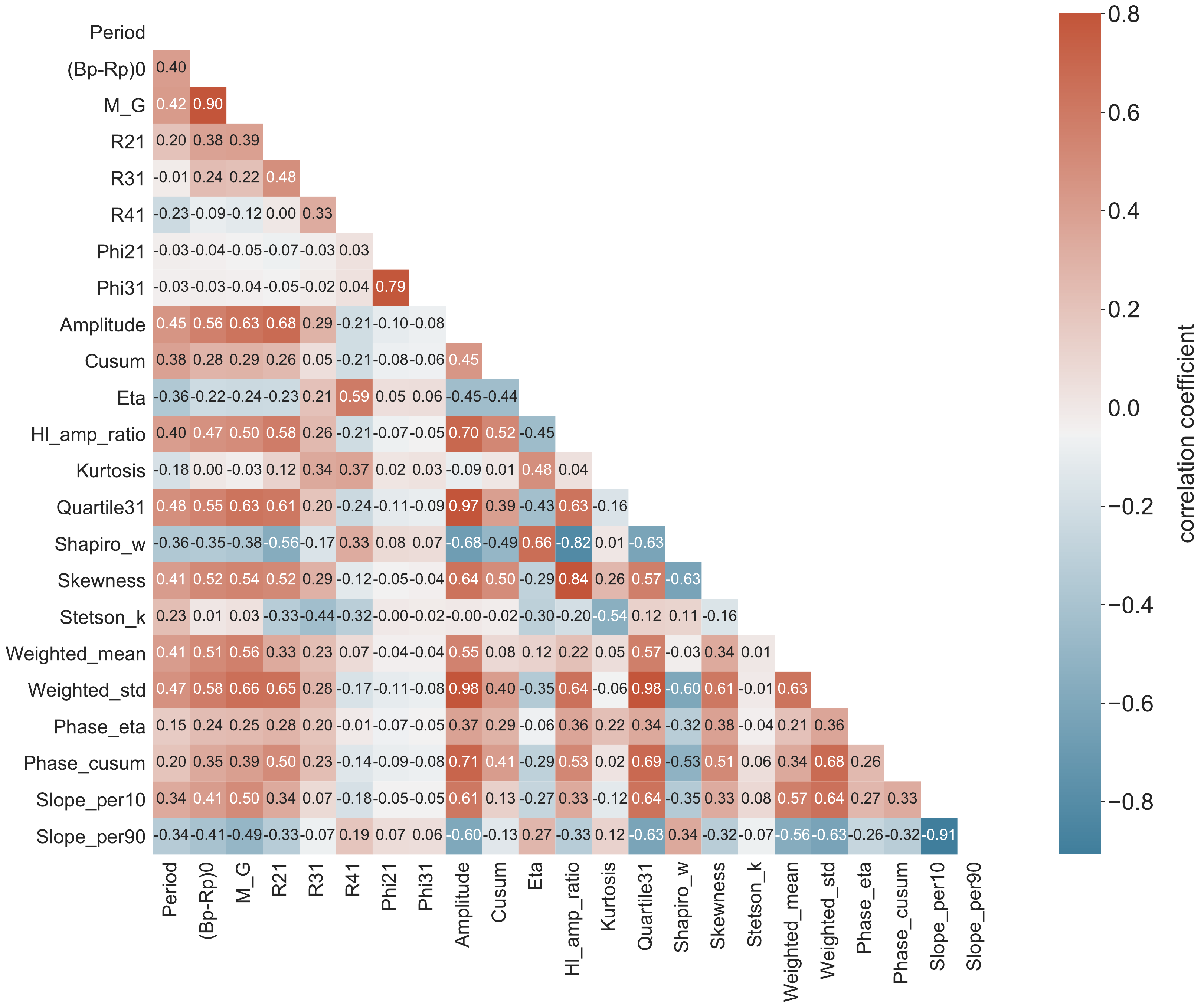}
    \caption{Correlation matrix of the features used in classification. The color of each square represents the value of the Pearson correlation coefficient between the corresponding features. A predominantly white color indicates a minimal correlation between a pair of features. Conversely, a thicker red (blue) color indicates a stronger positive (negative) correlation between the features.}
    \label{fig:correlation}
\end{figure*}

\begin{figure}
    \includegraphics[width=0.5\textwidth]{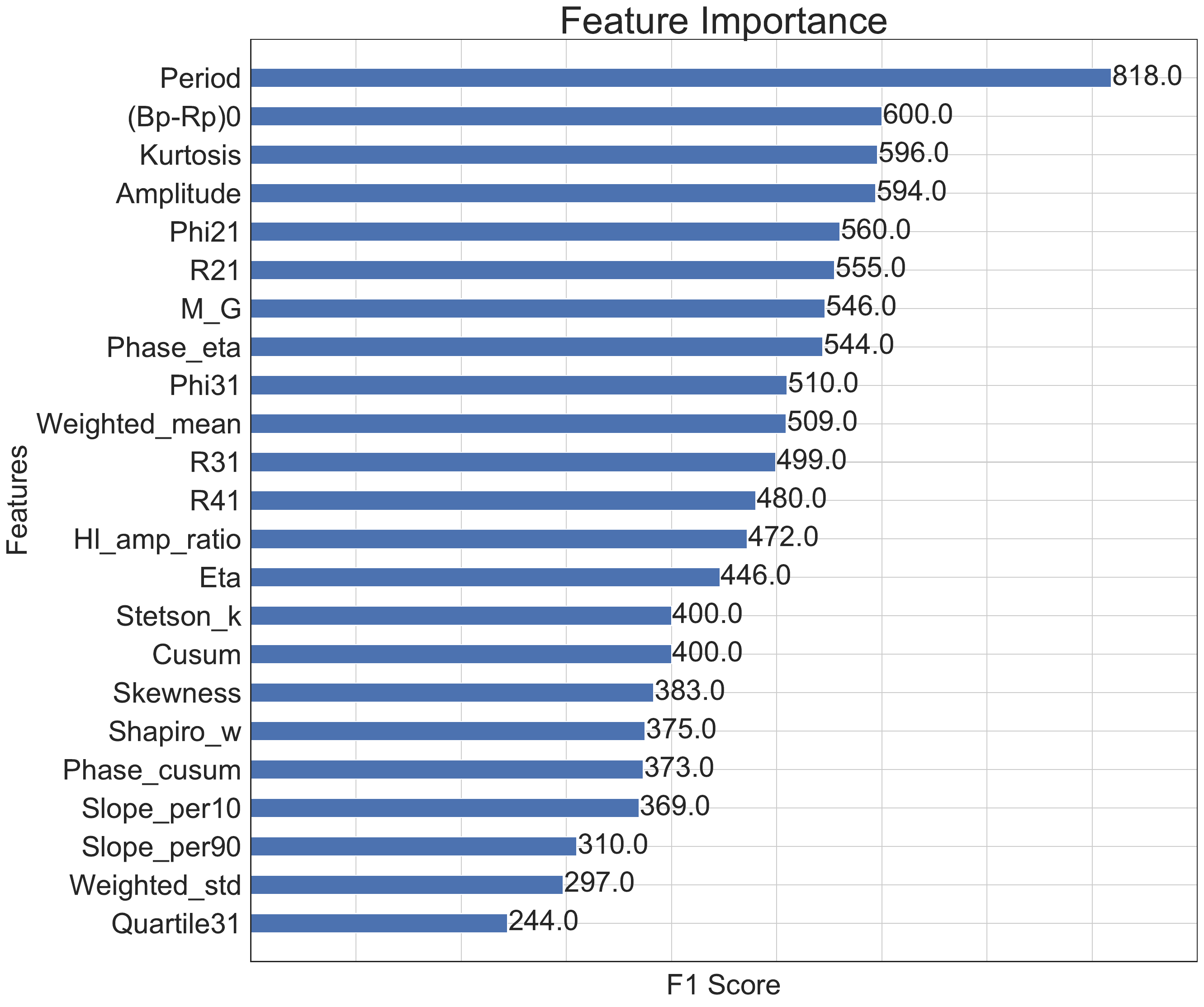}
    \caption{Feature importance of the feature set of the XGB classifier. The x-axis shows the F$_1$ scores (represents the gain) of each feature, and the y-axis displays the names of the features.}
    \label{fig:Feature importance}
\end{figure}

\begin{figure*}
    \centering
    \includegraphics[width=0.49\textwidth]{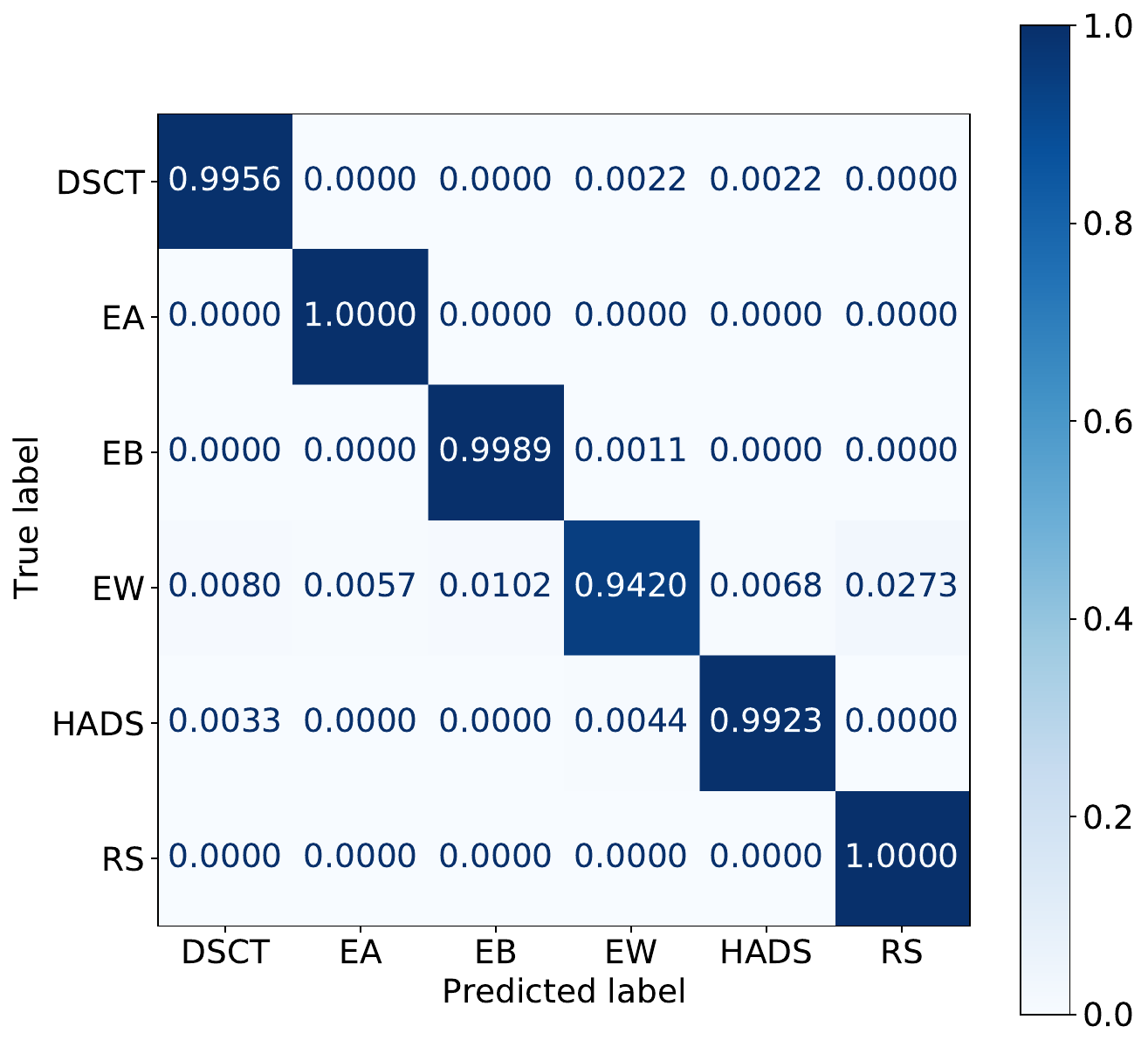}
    \includegraphics[width=0.49\textwidth]{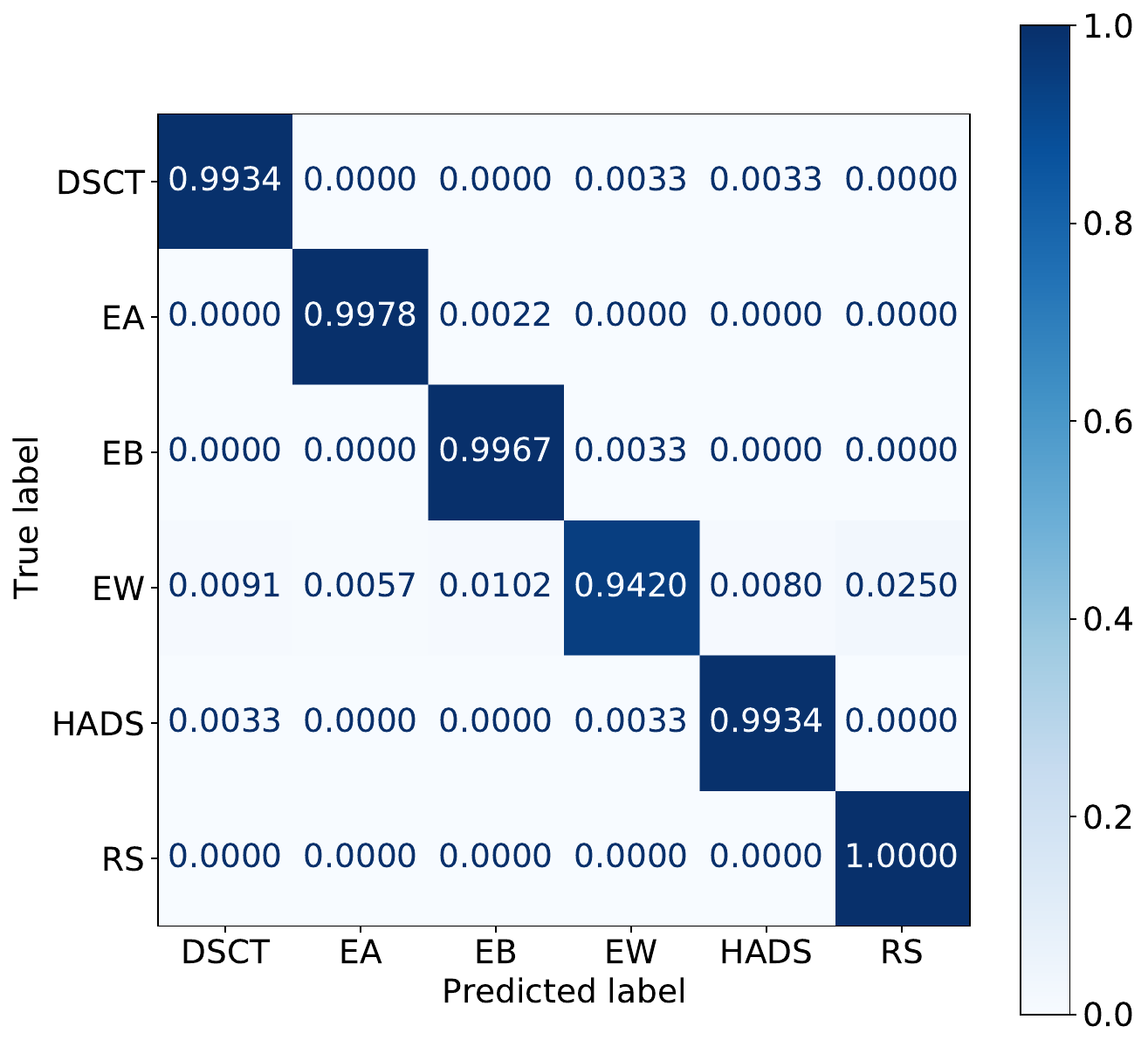}
    \caption{The normalized confusion matrix derived from the XGB classifier (left) and the RF classifier (right) based on the labelled dataset (Dataset-I). The x-axis represents the prediction class obtained from the classifier, and the y-axis displays the true group of a variable star.}
    \label{fig:confusion matrix}
\end{figure*}

In pursuit of higher prediction accuracy, we trained two classifiers, the XGBoost (XGB) classifier and the Random Forest (RF) classifier. Both classifiers would undergo training and testing in Dataset-I, and then predict the classes of each LCs in Dataset-II as well as providing a probability of the classification \citep{chen2015xgboost,breiman2001random}. XGBoost and random forest are two popular machine learning algorithms based on ensemble learning techniques, where multiple decision trees are integrated to generate predictions, constructing a powerful knowledge discovery and data mining model \citep{dong2020survey}. Both algorithms can be paralleled to speed up the training process and are robust to over-fitting. However, there are still certain differences between them. Random forest creates multiple weak decision trees independently, aggregating the predictions given by each tree to reach the ultimate result. XGBoost, using gradient boosting, optimizes the loss function by iteratively adding new trees. Comparing with neural networks (NNs), the two algorithms are more interpretable and require less preprocessing of the data. For instance, NNs exhibit sensitivity to the initial phases of LCs, which are external to the variability. Consequently, preprocessing steps or structural adjustments of the neural network are necessary \citep{zhang2021classification}. Moreover, XGBoost and random forest algorithms can provide insight into the importance of each feature input, allowing us to identify the key characteristics of the LCs to help distinguish different types. Although NNs may work better in large and complex datasets \citep{bishop1994neural}, the capabilities of our classifiers are sufficient for the study.

Dataset-I is extremely imbalanced, as illustrated in Figure \ref{fig:pie}, where the number of EWs is more than 70 times that of the least abundant EAs. Considering eclipsing binaries alone, VSX recorded 812 EAs, 121 EBs and 62,356 EWs with periods below 7.5 h, roughly consistent with the proportion of these variables in Dataset-I. Due to the small numbers of some classes and the modest size of Dataset-I, it is hard to build a balanced training set. Although XGBoost and Random Forest algorithms can handle imbalanced datasets, the result without any preprocessing was not satisfactory. So we used Synthetic Minority Oversampling Technique (SMOTE, \citealt{chawla2002smote}), an over-sampling technique, to prepare for a balanced training set. Based on the K-Nearest Neighbors algorithm, SMOTE synthesizes new instances for the minority classes by interpolating between existing samples from these classes, effectively increasing their representation. 

We extracted some basic features (i.e.,features not related to period and the light curve fitting) of the LCs with the help of the UPSILoN package \citep{kim2016package}. We simply used the ``extract\_features'' module instead of its classification algorithm. Due to the importance of period determination, we carefully calculated it through LSP (see also paper TMTS-II) and fit the LCs with a fourth-order Fourier function \citep{chen2020zwicky}:
\begin{equation}
    f=a_0 + \sum_{i=1}^{4} a_i\cos(2\pi it/P + \phi _i)
	\label{eq:fourier}
\end{equation}

where $a_i$ (i = 0,1,2,3,4) and $\phi _i$ (i = 1,2,3,4) represent the Fourier amplitudes and phases in each order. Notice that we used a fourth-order fitting, while UPSILoN chooses a third-order function. We studied the residual of the third-order fitting, finding that nearly half of the residuals exceed the error of the magnitude. On the contrary, the fourth-order fitting leads to only 7.4\% of the residuals above the magnitude error. Then we calculated some critical parameters like $R$$\mathrm{_{21}}$ and $\Phi$$\mathrm{_{21}}$ based on the fitting. Considering the challenge of classifying variable stars based on LCs alone, we included $Gaia$ $(B\mathrm{_p}-R\mathrm{_p})_0$ and $G$$\mathrm{_{abs}}$ into the feature list. Since only a small fraction of the sources in Dataset-I and II have LAMOST log $g$, $T$$\mathrm{_{eff}}$ and spectral type, we did not select these features. The complete set of features and their explanations in both classifiers are summarized in Table \ref{tab:features}.

Features that exhibit low correlations are generally preferred in machine learning classification, as they are more likely to provide independent information, thereby contributing to the model's ability to generalize and improve the classification accuracy. However, correlated features also have positive impacts. They enhance the stability of the algorithms by offering complementary information, and reduce the risk of overfitting by preventing the algorithms from relying excessively on an individual feature. Commonly, a correlation coefficient $r$ above 0.8 or below -0.8 would be considered a strong correlation. According to this criterion, as illustrated in Figure \ref{fig:correlation}, most of the features in our feature set exhibit weak or moderate correlations. However, strong correlations exist between certain features. For example, $G$$\mathrm{_{abs}}$ and $(B\mathrm{_p}-R\mathrm{_p})_0$ show a positive correlation with $r$ = 0.9, because of the distribution pattern in the CMD. Similarly, quartile$_{31}$, weighted\_std and amplitude show a strong correlation, because they represent different measurements of LC variabilities. Shapiro\_w and hl\_amp\_ratio exhibit a noticeable negative correlation with r = $-$0.82. This relationship can be readily comprehended, as shapiro\_w measures the extent to which a distribution deviates from a normal distribution. Larger deviations (as those found in the LCs of EAs) tend to result in an increase in hl\_amp\_ratio (a ratio between higher and lower amplitude relative to the average). 

Figure \ref{fig:Feature importance} shows the importance of all features used in our classification, where the importance is represented by a parameter named as F$_1$ score. The more a feature contributes to the improvement of classification accuracy, the greater its gain and therefore its F$_1$ score. Among various features, the most important one is the period, highlighting the critical role of precise period determination. Features that characterize the shape of the LCs from different aspects, such as kurtosis, amplitude and R$_{21}$, are of significant importance. Additionally, in accordance with our predictions, $Gaia$ $G$$\mathrm{_{abs}}$ and $(B\mathrm{_p}-R\mathrm{_p})_0$ also have great effects on the classifications. 

\subsection{Accuracy}

\begin{table}
	\centering
	\caption{The precision, recall and F$_1$ scores of the XGB classifier.}
	\label{tab:report@XGB}
	\begin{tabular}{cccc} 
		\hline
            \hline
		Type & Precision & Recall & F$_1$ score \\
		\hline   
            DSCT & 0.9956 & 0.9890 & 0.9923\\
            EA & 1.0000 & 0.9945 & 0.9972\\
            EB & 0.9989 & 0.9903 & 0.9946\\ 
            EW & 0.9420 & 0.9916 & 0.9662\\
            HADS & 0.9923 & 0.9912 & 0.9917\\
            RS & 1.0000 & 0.9729 & 0.9863\\  
            Macro-average & 0.9881 & 0.9883 & 0.9881\\ 
            Weighted-average & 0.9887 & 0.9883 & 0.9884\\
		\hline
	\end{tabular}
\end{table}

\begin{table}
	\centering
	\caption{The precision, recall and F$_1$ scores of the RF classifier.}
	\label{tab:report@RFC}
	\begin{tabular}{cccc} 
		\hline
            \hline
		Type & Precision & Recall & F$_1$ score \\
		\hline   
            DSCT & 0.9934 & 0.9879 & 0.9907\\
            EA & 0.9978 & 0.9944 & 0.9961\\
            EB & 0.9967 & 0.9882 & 0.9924\\
            EW & 0.9420 & 0.9893 & 0.9651\\
            HADS & 0.9934 & 0.9890 & 0.9912\\
            RS & 1.0000 & 0.9751 & 0.9874\\
            Macro-average & 0.9872 & 0.9873 & 0.9871\\
            Weighted-average & 0.9878 & 0.9873 & 0.9874\\
		\hline
	\end{tabular}
\end{table}

Three scores are used to evaluate the performance of the classifiers. Precision measures the fraction of true positives relative to the predicted positive instances:
\begin{equation}
    \mathrm{Precision=\frac{True\  Positives}{True\  Positives+False\  Positives}}.
	\label{eq:precision}
\end{equation}

Recall refers to the proportion of true positives among actual positive instances:
\begin{equation}
    \mathrm{Recall=\frac{True\  Positives}{True\  Positives + False\  Negatives}}.
	\label{eq:recall}
\end{equation}

 Where the true positives refer to the samples correctly classified as belonging to a specific class, while false positives indicate samples that do not belong to this class but are incorrectly classified as part of it. Moreover, false negatives represent the samples that actually belong to that class but are incorrectly classified as not belonging. A clear trade-off exists between these two measurements. Aiming to improve precision, the algorithm tends to minimize predicting positive samples to mitigate false positives. On the contrary, for optimizing recall, the model is inclined to select more positive samples to avoid missing any possible positives. Thus, F$_1$ score is the harmonic mean of precision and recall, which provides a balanced measure of the algorithm's performance:
\begin{equation}
    \mathrm{F_1\  score=\frac{2}{1/precision + 1/recall}}.
	\label{eq:f1-score}
\end{equation}

The normalized confusion matrices of both classifiers is shown in Figure \ref{fig:confusion matrix}. We used two measurements to evaluate the overall performance of the classifiers. Macro-average treats each class fairly, while weighted-average gives each type a weight that is proportional to the number of samples in that class. As our dataset is heavily imbalanced, weighted-average is more appropriate in measuring the performance of the classifiers. Some analogous patterns can be discerned in both confusion matrices. Both classifiers exhibit excellent performance in distinguishing EAs and RS CVn candidates, potentially owing to the distinctive shapes of their light curves. The overall performance of the XGB classifier is slightly better than the RF classifier. The macro-average and weighted-average F$_1$ scores of the XGB classifier are 98.81\% and 98.84\%, respectively. The precision, recall and F$_1$ scores for each type are summarized in Table \ref{tab:report@XGB}. The F$_1$ scores for all six types are higher than 96\%, which indicates that the variable stars are well distinguished by the XGB classifier. With F$_1$ scores $\geq$ 99\%, the DSCT, EA, EB and HADS classes are very well classified, while the scores of EWs appear to be relatively diminished.

For the RF classifier, its performance is summarized in Table \ref{tab:report@RFC}. All the F$_1$ scores are higher than 96\%, with the macro-average and weighted-average F$_1$ scores of 98.71\% and 98.74\%. Similar to the situation of the XGB classifier, the EW class has lower F$_1$ scores than other classes. In both classifiers, the relatively lower F$_1$ scores of EWs may potentially attributed to the contamination of EBs and RS CVn candidates, as illustrated in Figure \ref{fig:confusion matrix}. EBs are semi-detached eclipsing binaries occupying an intermediate evolutionary state between EAs and EWs, and in some cases their LCs closely resemble those of EWs, leading to challenges in their differentiation. In fact, sometimes the classification of eclipsing binaries has been simplified to primarily two categories \citep{chen2020zwicky}: EA-type (detached) and EW-type (contact). Furthermore, some RS CVn variables are themselves eclipsing binaries, which explains the occasional similarity between their LCs and those of EWs.

In both classifiers, the F$_1$ score for DSCT, HADS, EA and EB exceeds 99\%, while EW slightly lags behind at 96\%. These results underscore the effectiveness of our feature selection and algorithm training processes. However, we shall point out that the confusion matrices and the scores are derived from Dataset-I (the labelled dataset), while Dataset-II (the unlabelled dataset) might contain fainter objects with noisier light curves, making them more susceptible to misclassification. As a consequence, the scores might be degraded in Dataset-II. Additionally, we acknowledge that our training and test datasets are relatively limited in size and may not yield the same level of accuracy if expanded. 

When appling the classifiers to Dataset-II, we find that some objects do not belong to the 6 main classes explored in this paper. To avoid misclassifications, we conducted a comprehensive manual inspection of all sources, which should minimize the probability of such errors.

\section{Catalogues} 
\label{sec:cata}

\begin{figure*}
    \centering
    \includegraphics[width=1\textwidth]{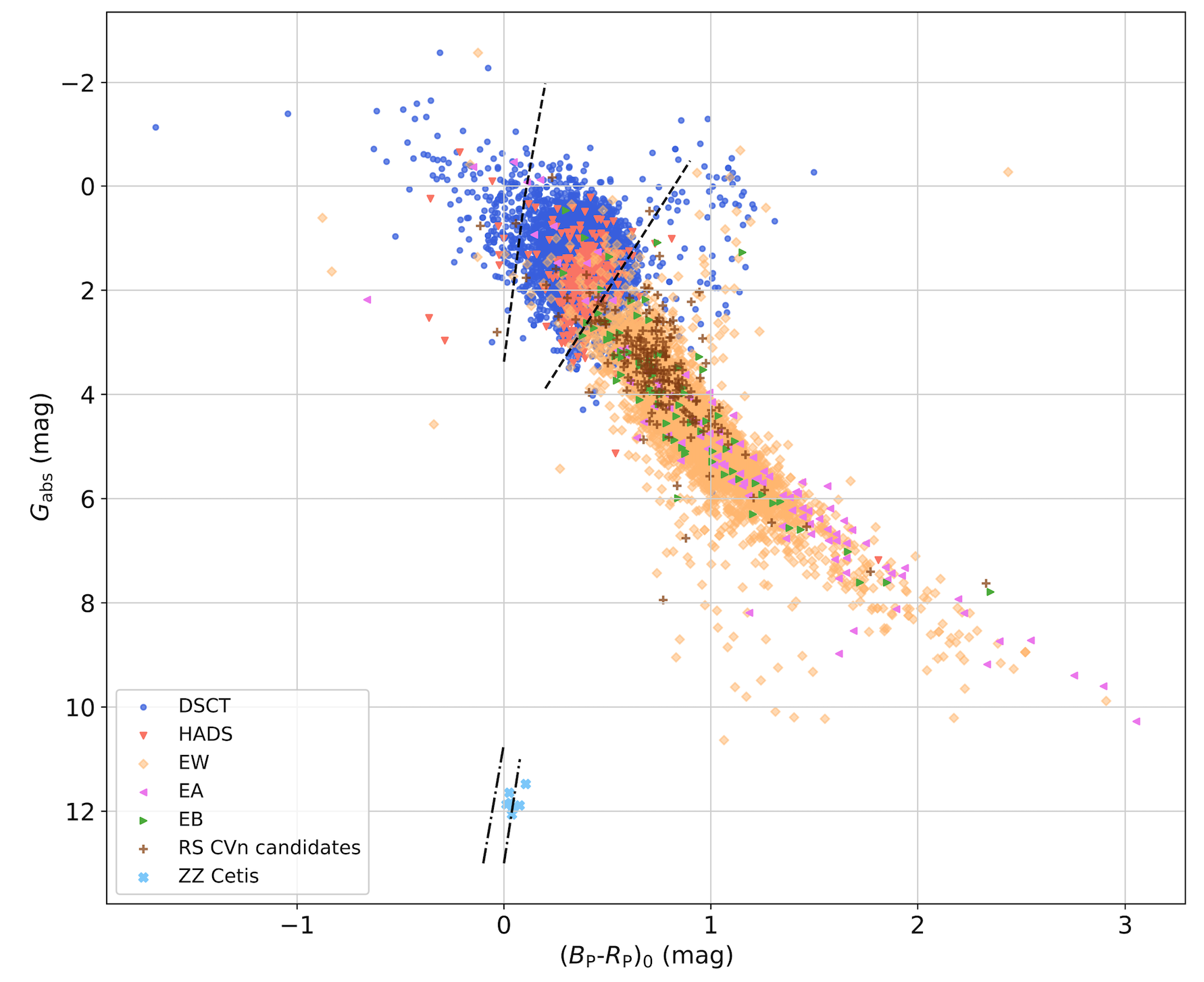}
    \caption{Distribution of variable stars in TMTS Catalog of Periodic Variable Stars across the color-magnitude diagram. Color-coded symbols represent different classes of variable stars, include typical $\delta$ Scuti stars (DSCT, royal blue), high amplitude $\delta$ Scuti stars (HADS, red), EW-type eclipsing binaries  (EW, orange), EA-type eclipsing binaries (EA, pink), EB-type eclipsing binaries (EB, green), RS CVn candidates (RS, brown) and ZZ Ceti variables (ZZ, azure). The dashed lines and dash-dotted lines indicate the instability strip edges for $\delta$ Scuti stars \citep{murphy2019gaia} and ZZ Ceti stars \citep{caiazzo2021highly}, respectively. The edges of the instability strip are calculated by using the relationship of $Gaia$ $B_\mathrm{P}-R_\mathrm{P}$ and $T_\mathrm{eff}$ (see also \citealt{jordi2010gaia}).}
    \label{fig:CMD}
\end{figure*}

\begin{figure*}
    \centering
    \includegraphics[width=1\textwidth]{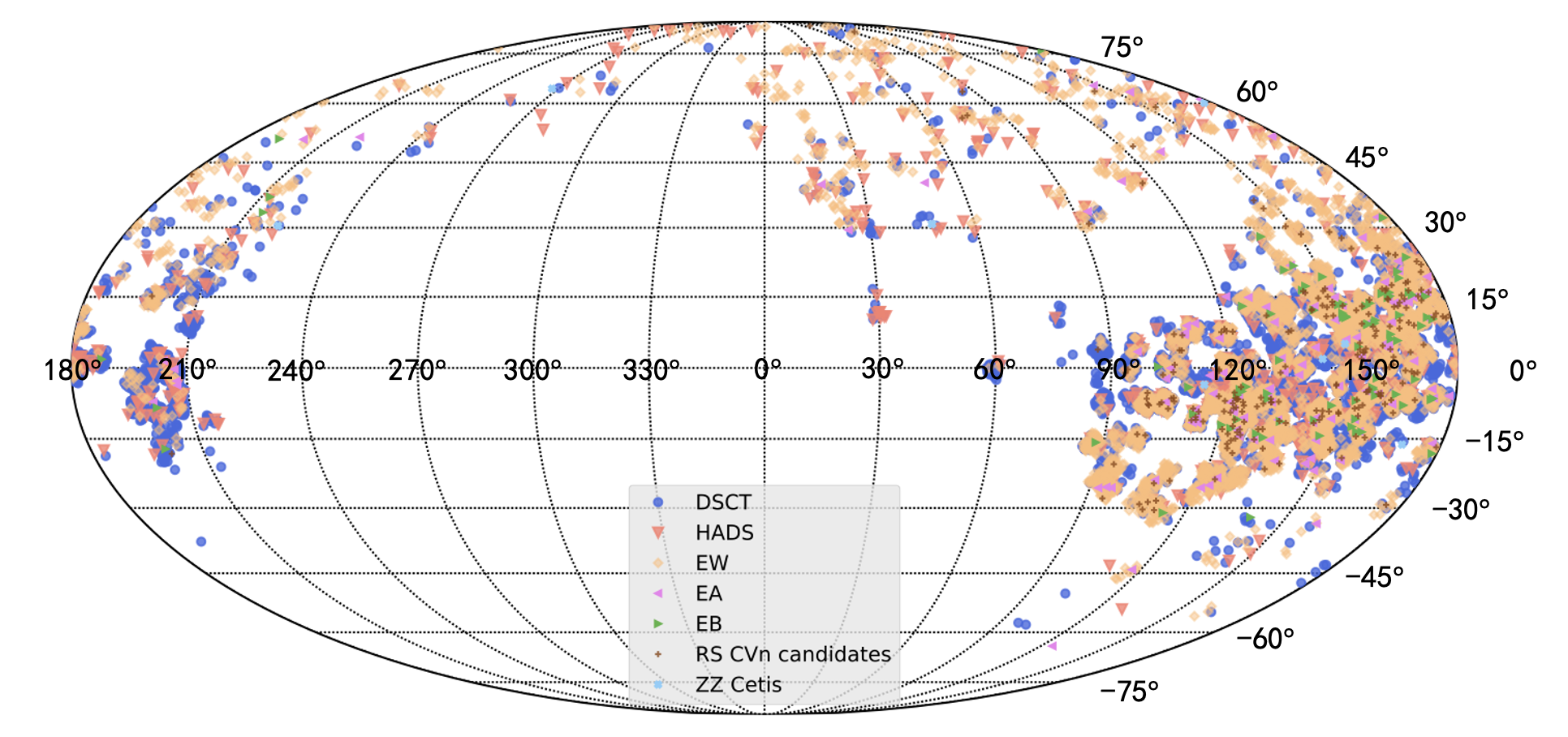}
    \caption{Distribution of variable stars in TMTS Catalog of Periodic Variable Stars in Galactic coordinates.}
    \label{fig:Distribution}
\end{figure*}

\begin{table}
	\centering
	\caption{All and newly discovered variables in TMTS Catalogues of Periodic Variable
Stars.}
	\label{tab:classification}
	\begin{tabular}{ccc} 
		\hline
            \hline
		Type & Total number	& Newly discovered (Fraction) \\
		\hline
            DSCT & 4876	& 4297 (88.13\%) \\		
            EA & 117 & 18 (15.38\%) \\
            EB & 84 & 18 (21.43\%) \\
            EW & 5698 & 780 (13.69\%) \\
            HADS & 628 & 74 (11.78\%) \\
            RS CVn & 226 & 108 (47.79\%) \\
            ZZ Ceti & 9 & 6 (66.67\%) \\
		\hline
	\end{tabular}
\end{table}

\begin{sidewaystable*}
    \centering
    \caption{Example catalogue for typical $\delta$ Scuti stars (DSCT) and high amplitude $\delta$ Scuti stars (HADS) in TMTS Catalogues of Periodic Variable Stars.}
    \label{tab:dsct_hads}
    \begin{tabular}{cccccccccccccc} 
        \hline
        \hline
        Type & source I.D. & R.A. & Dec. & Period & Amp & $R_{21}$ & $\Phi_{21}$ & $G\mathrm{_{abs}}$ & $(B\mathrm{_p}-R\mathrm{_p})_0$ & spec.type & VSX type & $T\mathrm{_{eff}}$ & log g\\
         & & deg & deg & hours & mag & & & mag & mag & & & K & cm$^{-2}$\\
        \hline
        HADS & TMTS J00000450+5146157 & 0.01875	& 51.77104 & 2.28702 & 0.077 & 0.464 & 0.435 & & & & DSCT\\
        DSCT & TMTS J00001301+5520331 & 0.05422 & 55.34254 & 2.39646 & 0.018 & 0.028 & 0.701 &	1.086 & 0.412\\ 
        DSCT & TMTS J00001318+5144112 & 0.05494 & 51.73646 & 1.64768 & 0.036 & 0.262 &	-0.180 & 2.428 & 0.356 & & DSCT\\
        DSCT & TMTS J00003793+5423372 & 0.15805 & 54.39367 & 2.22930 &	0.033 &	0.384 &	0.584 &	1.206 &	0.375\\ 
        DSCT & TMTS J00004133+6034008 & 0.17223 & 60.56690 & 0.74820 &	0.006 &	0.069 &	0.288 &	1.277 &	0.150\\ 
        DSCT & TMTS J00004531+5857086 & 0.18880 & 58.95239 & 1.27363 &	0.010 &	0.113 &	0.108 &	1.494 &	0.345 &	F0 & DSCT & 7148.6 & 3.913\\ 
        DSCT & TMTS J00004603+5649520 & 0.19179 & 56.83111 & 1.72859 &	0.012 &	0.179 &	1.436 &	1.978 &	0.406 &	F0 & & 7329.5 & 3.951\\ 
        DSCT & TMTS J00011153+5402497 & 0.29804 & 54.04713 & 0.86593 &	0.016 &	0.022 &	-0.003 & 1.565 & 0.271\\		
        HADS & TMTS J00011212+5756134 & 0.30052	& 57.93705 & 2.39443 & 0.179 & 0.165 &	0.142 & & & & HADS\\
        DSCT & TMTS J00011472+3332389 & 0.31135 & 33.54413 & 1.05077 &	0.012 &	0.022 &	-0.470 & 1.659 &	0.305\\		
        HADS & TMTS J00011653+5430249 & 0.31889	& 54.50691 & 1.09143 & 0.057 & 0.202 & -0.768 & 5.130 & 0.539	& & HADS\\
        HADS & TMTS J00013456+6012530 &	0.39398	& 60.21473 & 2.72071 & 0.092 &	0.208 & -2.888 & & & & DSCT\\
        DSCT & TMTS J00014233+6725590 & 0.42638 & 67.43332 & 2.56480 &	0.016 &	0.205 &	0.872 & & & & DSCT\\		
        DSCT & TMTS J00014879+5616112 & 0.45328 & 56.26978 & 2.57181 &	0.012 &	0.118 &	-0.686 & 1.393 & 0.473\\ 	
        HADS & TMTS J00020221+5602430 &	0.50922 & 56.04529 & 1.52818 & 0.208 & 0.188 & 0.768 & & & & HADS\\
        DSCT & TMTS J00020700+5317162 & 0.52919 & 53.28783 & 1.77589 & 0.016 & 0.222 &	1.065 &	1.241 &	0.352\\
        HADS & TMTS J00021894+5353057 &	0.57894 & 53.88492 & 2.91048 & 0.094 & 0.141 & -0.116 & 0.981 & 0.443 & & HADS\\
        DSCT & TMTS J00023993+5224248 & 0.66637 & 52.40689 & 0.70615 & 0.008 &	0.093 &	-0.698 & 1.791 & 0.366\\ 		
        DSCT & TMTS J00030625+4927079 & 0.77606 & 49.45219 & 0.99118 & 0.007 &	0.229 &	-0.280 & 1.669 &	0.356 &	F0 & & 7359.6 & 3.902\\
        DSCT & TMTS J00030676+6111553 & 0.77816 & 61.19869 & 1.66266 & 0.013 &	0.058 &	-0.463 & 0.497 & 0.259\\ 		
        DSCT & TMTS J00031014+7056063 & 0.79226 & 70.93509 & 1.73341 & 0.008 &	0.154 &	-0.553 & 0.810 &	0.396 & & DSCT\\		
        DSCT & TMTS J00031753+5548560 & 0.82304 & 55.81556 & 0.87144 & 0.008 &	0.143 &	0.012 &	1.837 &	0.321\\
        DSCT & TMTS J00032377+6015228 & 0.84906 & 60.25634 & 0.37645 & 0.005 &	0.155 &	1.833 &	2.150 &	0.141\\
        DSCT & TMTS J00032731+5057573 & 0.86380 & 50.96592 & 2.19325 & 0.008 &	0.137 &	-0.230 & 2.129 &	0.435\\
        HADS & TMTS J00051479+3233034 & 1.31163	& 32.55094 & 1.36111 & 0.240 & 0.307 & -0.745	& & & &	HADS\\
        HADS & TMTS J00085079+1831103 &	2.21162	& 18.51952 & 1.51535 & 0.081 & 0.147 & 0.122 & & & & DSCT\\			
        HADS & TMTS J00103232+3223484 &	2.63466	& 32.39677 & 1.76144 & 0.059 & 0.175 & 0.242 & 1.903 & 0.352	& A7V & HADS(B)\\		
        HADS & TMTS J00145409+7053126 &	3.72537	& 70.88682 & 1.85296 & 0.146 & 0.203 & 1.032	& &	& & HADS\\	
        HADS & TMTS J00165257+6317533 &	4.21905	& 63.29815 & 2.59283 & 0.190 & 0.377 & -0.656 & 2.350 & 0.474	& &	HADS\\	
        HADS & TMTS J00172714+5825592 &	4.36309	& 58.43312 & 2.19222 & 0.147 & 0.203 & -0.101 & 0.743 & 0.231	& & HADS\\			
        \hline
    \end{tabular}  
\end{sidewaystable*}

\begin{sidewaystable*}
    \centering
    \caption{Example catalogue for EA, EB and EW-type binaries in TMTS Catalogues of Periodic Variable Stars.}
    \label{tab:eclipsing}
    \begin{tabular}{cccccccccccccc} 
        \hline
        \hline
        Type & source I.D. & R.A. & Dec. & Period & Amp & $R_{21}$ & $\Phi_{21}$ & $G\mathrm{_{abs}}$ & $(B\mathrm{_p}-R\mathrm{_p})_0$ & spec.type & VSX type & $T\mathrm{_{eff}}$ & log g\\
         & & deg & deg & hours & mag & & & mag & mag & & & K & cm$^{-2}$\\
        \hline
        EW & TMTS J00000019+3208472	& 0.00079 & 32.14645 & 3.40593 & 0.097 & 0.040 & -1.045 & 5.051 & 1.032 & G9 & EW & 5011.9 & 4.102\\ 
        EW & TMTS J00000026+3112061	& 0.00109 & 31.20170 & 4.29466 & 0.134 & 0.088 & 0.785 & 3.925 & 0.767 & & EW\\	
        EW & TMTS J00000325+6922130	& 0.01354 & 69.37055 & 4.47799 & 0.172 & 0.192 & -0.016 & 2.943 & 0.494 & & EW\\	
        EW & TMTS J00000435+5105040	& 0.01812 & 51.08472 & 4.11442 & 0.089 & 0.073 & -0.886 & & & & EW\\	
        EW & TMTS J00000593+5559002	& 0.02473 &	55.98338 & 5.02115 & 0.127 & 0.131 & -1.353 & & & F0 & EW\\ 
        EB & TMTS J00001027+4951403	& 0.04278 &	49.86118 & 3.96473 & 0.193 & 0.229 & 0.077 & 4.318 & 0.850 & F0 & EW\\ 
        EW & TMTS J00002303+5424323	& 0.09595 & 54.40896 & 4.98939 & 0.085 & 0.069 & -1.011 & 3.340 & 0.766 & & EW\\	
        EW & TMTS J00002623+5246201	& 0.10928 & 52.77224 & 3.99245 & 0.293 & 0.316 & -0.997 & 4.301 & 0.825 & & EW\\
        EW & TMTS J00003480+5349116	& 0.14498 & 53.81989 & 2.75730 & 0.204 & 0.248 & 0.008 & 6.322 & 1.399 & & EW\\	
        EW & TMTS J00004595+5623556	& 0.19146 & 56.39879 & 4.39351 & 0.284 & 0.359 & -0.247 & 4.382 & 0.940 & & EW\\	
        EW & TMTS J00004719+7117019	& 0.19663 & 71.28387 & 3.83828 & 0.114 & 0.161 & 1.002 & 4.249 & 0.864 & & EW\\	
        EW & TMTS J00011366+3205104	& 0.30690 & 32.08621 & 2.90506 & 0.038 & 0.071 & 0.871 & 3.655 & 0.769\\			
        EW & TMTS J00011386+3206380	& 0.30777 & 32.11056 & 3.65823 & 0.246 & 0.307 & -0.924 & & & & EW\\	
        EW & TMTS J00011485+5514513	& 0.31188 & 55.24759 & 4.79292 & 0.052 & 0.047 & 1.000 & 3.327 & 0.611 & & EW\\	
        EW & TMTS J00011492+4924399	& 0.31215 & 49.41109 & 3.65327 & 0.135 & 0.117 & -0.252 & & & & EW\\
        EA & TMTS J00030963+6843573	& 0.79011 & 68.73259 & 4.30887 & 0.030 & 0.390 & -1.082 & 8.538 & 1.691 & & EA\\	
        EA & TMTS J00070364+5654212	& 1.76518 & 56.90590 & 4.39351 & 0.168 & 0.421 & -0.161 & & & K3 & EA & 4928.0 & 4.296\\ 
        EB & TMTS J00144055+3109196	& 3.66896 & 31.15544 & 1.44881 & 0.009 & 0.190 & -0.663 & 0.991 & 0.391 & & EB\\	
        EA & TMTS J00193116+6136261	& 4.87983 & 61.60724 & 2.48298 & 0.043 & 0.109 & -0.098 & -0.461 & 0.048 & & EA\\	
        EB & TMTS J00342250+5044134	& 8.59374 & 50.73704 & 4.70277 & 0.041 & 0.155 & -1.386 & 2.206 & 0.615\\ 			
        EA & TMTS J00342377+1832050	& 8.59902 & 18.53499 & 3.73003 & 0.105 & 0.440 & -0.710 & 8.121 & 1.897 & & EA\\	
        EA & TMTS J00411474+3717343	& 10.31142 & 37.29286 & 3.21250 & 0.093 & 0.372 & -0.961 & 5.938 & 1.349\\ 		
        EA & TMTS J00412268+5538134	& 10.34451 & 55.63706 & 3.37239 & 0.182 & 0.555 & -0.902 & 4.833 & 0.643\\			
        EA & TMTS J00421197+7755577	& 10.54989 & 77.93268 & 1.12945 & 0.013 & 0.268 & -0.541 & 2.181 & 0.455 & & EA\\	
        EA & TMTS J00495796+5337533	& 12.49148 & 53.63147 & 2.22108 & 0.036 & 0.614 & -0.065 & 3.758 & 0.739\\
        EA & TMTS J00501746+3753114	& 12.57275 & 37.88651 & 3.73250 & 0.342 & 0.588 & -0.045 & 6.587 & 1.565  & K7 & EA & 3888.1 & 4.237\\
        EA & TMTS J00502180+5403015	& 12.59084 & 54.05041 & 5.18251 & 0.048 & 0.722 & -2.997 & 3.156 & 0.549\\
        EB & TMTS J00505429+5005103	& 12.72620 & 50.08618 & 2.94622 & 0.186 & 0.317 & 0.007 & 6.596 & 1.434\\
        EB & TMTS J00513823+5706077	& 12.90928 & 57.10213 & 4.98543 & 0.071 & 0.551 & -0.868 & 5.992 & 0.842 & & EB:\\	
        EA & TMTS J00550688-0057031	& 13.77865 & -0.95087 & 2.85317 & 0.161 & 0.446 & -1.432 & 9.395 & 2.754 & & EA\\	
        \hline
        \end{tabular}
\end{sidewaystable*}

\begin{sidewaystable}
    \centering
    \caption{Example catalogue for candidates of RS CVn in TMTS Catalogues of Periodic Variable Stars.}
    \label{tab:rs}
    \begin{tabular}{cccccccccccccc} 
        \hline
        \hline
        Type & source I.D. & R.A. & Dec. & Period & Amp & $R_{21}$ & $\Phi_{21}$ & $G\mathrm{_{abs}}$ & $(B\mathrm{_p}-R\mathrm{_p})_0$ & spec.type & VSX type & $T\mathrm{_{eff}}$ & log g\\
         & & deg & deg & hours & mag & & & mag & mag & & & K & cm$^{-2}$\\
        \hline
        RS & TMTS J00023663+3115152 & 0.65262 & 31.25423 & 4.48988 & 0.042 & 0.055 & -0.474 & 4.051 & 0.895 & G9 & & 5637.3 & 4.20814\\
        RS & TMTS J00031400+5051010 & 0.80835 & 50.85055 & 5.15414 & 0.032 & 0.167 & 0.151 & 4.822 & 0.798	& &	& & \\
        RS & TMTS J00074144+7004396 & 1.92266 & 70.07766 & 5.11770 & 0.032 & 0.132 & -0.064 & 2.899 & 0.596 & & RS: & & \\		
        RS & TMTS J00084725+6206404 & 2.19689 & 62.11123 & 4.27073 & 0.037 & 0.059 & 0.645 & 2.089 & 0.743 & &	& & \\
        RS & TMTS J00090631+6147561 & 2.27630 & 61.79891 & 5.08420 & 0.088 & 0.065 & 0.431 & 2.595 & 0.739 & & RS &  & \\ 	
        RS & TMTS J00170757+7051209 & 4.28156 & 70.85579 & 4.88508 & 0.037 & 0.223 & 0.425 & 4.332 & 0.878 & &	RS & & \\	
        RS & TMTS J00175991+5635079 & 4.49963 & 56.58553 & 3.64480 & 0.115 & 0.057 & 0.904 & 6.461 & 1.294 & &	RS: & & \\
        RS & TMTS J00230426+7028187 & 5.76774 & 70.47185 & 3.48299 & 0.050 & 0.167 & -0.547 & 3.319 & 0.650 & & RS:	& & \\
        RS & TMTS J00305793+5548374 & 7.74138 & 55.81038 & 3.99691 & 0.038 & 0.411 & -0.705 & 2.529 & 0.749 & & & & \\
        RS & TMTS J00331241+5337429 & 8.30172 & 53.62857 & 4.03805 & 0.016 & 0.136 & -1.086 & 4.710 & 0.965 & G7 &	& 5379.0 & 4.095\\	
        \hline
        \end{tabular}
\end{sidewaystable}

\begin{sidewaystable}
    \centering
    \caption{Catalogue for ZZ Ceti stars in TMTS Catalogues of Periodic Variable Stars.}
    \label{tab:zz}
    \begin{tabular}{cccccccccccccc} 
        \hline
        \hline
        Type & source I.D. & R.A. & Dec. & Period & Amp & $R_{21}$ & $\Phi_{21}$ & $G\mathrm{_{abs}}$ & $(B\mathrm{_p}-R\mathrm{_p})_0$ & spec.type & VSX type & $T\mathrm{_{eff}}$ & log g\\
         & & deg & deg & hours & mag & & & mag & mag & & & K & cm$^{-2}$\\
        \hline
        ZZ & TMTS J03453796+5707553 & 56.40817 & 57.13203 & 0.22635 & 0.055 & 0.095 & 0.256 & 11.477 & 0.105\\
        ZZ & TMTS J04185665+2717472 & 64.73606 & 27.29645 & 0.13766 & 0.025 & 0.181 & -0.865 & 11.645 & 0.025 & WD	& ZZA\\
        ZZ & TMTS J04332464+5520114 & 68.35266 & 55.33651 & 4.76767 & 0.107 & 0.313 & 0.088 & 12.062 & 0.039 & WD\\
        ZZ & TMTS J08550719+0635392 & 133.77994 & 6.594231 & 0.21981 & 0.058 & 0.267 & -1.361 & 11.888 & 0.076	& & ZZA\\
        ZZ & TMTS J10423368+4057149 & 160.64034 & 40.95413 & 0.25441 & 0.035 & 0.378 & 0.164 & 11.840 & 0.032 & WD	& ZZA\\
        ZZ & TMTS J11515421+0528344 & 177.97587 & 5.47621 & 0.31778 & 0.0112 & 0.246 & -0.009 & 11.951 & 0.041\\
        ZZ & TMTS J17184056+2524311 & 259.66900 & 25.40863 & 0.13813 & 0.033 & 0.167 & -0.156 & 11.870 & 0.012\\
        ZZ & TMTS J17184064+2524314 & 259.66932 & 25.40873 & 0.19933 & 0.079 & 0.155 & 0.033 & 11.870 & 0.012\\
        ZZ & TMTS J23450729+5813146 & 356.28038 & 58.22072 & 0.26931 & 0.080 & 0.183 & -0.674 & 11.948 & 0.046\\
        \hline
        \end{tabular}
\end{sidewaystable}

\begin{table*} 
	\centering
	\caption{Catalogue for candidates of SX Phe stars in TMTS Catalogues of Periodic Variable Stars.}
	\label{tab:SX Phe}
	\begin{tabular}{cccccccccc} 
		\hline
            \hline
	    source I.D. & l & b & Period & [Fe/H] & spec.type & $T\mathrm{_{eff}}$ & log g & $G\mathrm{_{abs}}$ & $(B\mathrm{_p}-R\mathrm{_p})_0$ \\
             & deg & deg & hours & & & K & cm$^{-2}$ & mag & mag\\
		\hline   
            TMTS J02063796+2208355 & 145.02592 & -37.48132 & 1.35 & -1.287 & A3IV & 7149.580 & 4.349 & 3.086 & 0.340\\
            TMTS J08454910+1236160 & 214.18678 & 31.01094 & 1.47 & -1.076 & A5V & 7203.620	& 4.325 & 2.414	& 0.428\\
            TMTS J09360476+4440412 & 175.32518 & 47.48746 & 1.30 & -1.278 & A6IV & 7145.050 & 4.259 & 2.525 & 0.293\\
            TMTS J11440047+3629278 & 175.50250 & 72.92707 & 0.98 & -1.113 & A3IV & 7311.230 & 4.310 & 1.999 & 0.347\\
            TMTS J13122682+3112462 & 75.87687 & 83.86032 & 0.96 & -1.145 & A3IV & 7148.841 & 4.309 & 2.093	& 0.354\\
            TMTS J14551976+3950046 & 67.26214 & 61.40951 & 1.09 & -1.411 & A1V & 7118.290 & 4.386 & 2.779 & 0.335\\
            TMTS J17211940+2608082 & 48.76646 & 30.46348 & 1.47 & -1.184 & A3IV & 7116.610 & 4.237 & 2.297 & 0.281\\
            TMTS J23534614+2957391 & 108.31427 & -31.26870 & 1.12 & -1.044 & A5V & 7328.820 & 4.451 & 2.948 & 0.320\\     
		\hline
	\end{tabular}
\end{table*}

We identified 11,638 variables stars based on the first three-year surveys, which we regard as TMTS Catalogues of Periodic Variable Stars, including 4876 DSCT (including 910 released in the short-period variable stars catalog in TMTS-II), 628 HADS (including 166 released in the short-period variable stars catalog in paper TMTS-II), 117 EA-type eclipsing binaries, 84 EB-type eclipsing binaries, 5698 EW-type eclipsing binaries, 226 candidates of RS CVn variables and 9 ZZ Ceti stars. Note that the (910+166=)1076 $\delta$ Scuti stars identified in TMTS-II were all classified as $\delta$ Scuti stars, as TMTS-II did not distinguish between DSCT and HADS. All the LCs in our catalog have been visually checked. Table \ref{tab:classification} shows all and newly discovered variable stars we identified. Among them, 4297 DSCT are newly discovered, accounting for 88.13\% of all DSCT in our catalogue. Comparing to the 9162 DSCT recorded by VSX (including known and suspected DSCT, DSCTr and DSCTc), newly discovered DSCT from our catalogue greatly increases the number of known samples, demonstrating the ability of TMTS to discover short-period variables (see also \citealt{lin2023minute}). In comparison, other large-scale surveys are less efficient in identifying $\delta$ Scuti stars because of short periods intrinsic to these variables. In our catalog, the shortest period observed for $\delta$ Scuti stars is 0.21 h, with 0.35\% possessing periods below 0.5 hours. Furthermore, our survey exhibits higher fractions of $\delta$ Scuti stars with periods below 1 h and 1.5 h compared to other surveys, underscoring the TMTS's pronounced proficiency in detecting short-period variables.

Table \ref{tab:dsct_hads} shows an example of the DSCT and HADS in our catalogue, while Table \ref{tab:eclipsing} shows an example of EA-, EB- and EW-type eclipsing binaries. In Table \ref{tab:rs}, we summarize an example of RS CVn candidates, and we show the catalogue of  ZZ Ceti stars in Table \ref{tab:zz}. The machine-readable version of these catalogues can be found online. As the photometric period can be obtained directly from the LC and is used as a feature of the classifiers, the ``Period'' in our catalogue refers to the photometric period instead of the orbital period. For binaries, we can double the photometric period to obtain the orbital period. The catalogues contain the source ID, position (R.A. and decl.), period (photometric period for binaries), LC parameters (amplitude, R$_{21}$, $\Phi_{21}$), VSX type, as well as information from $Gaia$ DR2 (color and magnitude) and LAMOST DR7 (spectral type, effective temperature and log g). It is evident that fewer than 20\% of the variables in our catalogue possess LAMOST information, which imposes limitations on our ability to accurately classify certain subtypes.

We discovered 226 RS CVn candidates, among which 108 are newly identified. A notable characteristic of RS CVn is the occurrence of prominent Ca II H and K emission lines in the spectra, signifying increased chromospheric activity. As the confirmation of chromospheric activity requires additional information from the spectra \citep{drake2014catalina}, the variable stars in the class of ``RS'' are merely RS CVn candidates. 

In Figure \ref{fig:CMD} we show the distribution of the variable stars in our catalogues across the CMD. Consistent with previous studies \citep{eyer2019gaia}, $\delta$ Scuti stars locate in the lower part of the instability strip. We marked the red-edge and blue-edge of the instability strips of $\delta$ Scuti stars \citep{murphy2019gaia} and ZZ ceti stars \citep{caiazzo2021highly}. \citet{murphy2019gaia} defined the instability strip as specific boundaries in the CMD where about 20\% of the stars are pulsators. While the fraction of pulsators would tend to be lower outside the strip, this tendency does not necessarily preclude the existence of pulsators beyond the strip. Therefore, it is not unexpected that some $\delta$ Scuti stars are found outside the instability strip. Moreover, the effective temperatures of our $\delta$ Scuti sample are inferred from $Gaia$ $(B\mathrm{_p}-R\mathrm{_p})_0$ as in \citealt{jordi2010gaia}, which may also introduce  additional uncertainties.

With the help of the $Gaia$ CMD, we identified 9 ZZ Ceti stars, 5 of which have been reported in paper TMTS-II and \citealt{guo2023properties,guo2024variable} did detailed analysis of TMTS J23450729+5813146 and TMTS J17184064+2524314, respectively. ZZ Ceti stars are the most populous group of pulsating white dwarfs, with periods between 100 and 1200s. They occupy a distinct and narrow area in the instability strip of the CMD, rendering them more readily detectable \citep{fontaine2003confirmation}.

\subsection{SX Phe stars}

SX Phoenicis (SX Phe) stars are a subtype of $\delta$ Scuti stars. The primary criterion for distinguishing SX Phe stars from classical $\delta$ Scuti stars is their metal abundance. Most $\delta$ Scuti stars exhibit metal abundances similar to that of the Sun, placing them in the category of young, metal-rich population I stars typically found in the Galactic disk. In contrast, SX Phe stars belong to Population II stars. They possess relatively lower metal abundances, generally falling within the range of $-$1.5 $<$ [Fe/H] $<$ $-$1.0 \citep{mcnamara2011delta}, and are typically located in Galactic halos or globular clusters. Taking into account this distribution tendency, there is an enhanced likelihood of detecting SX Phe stars in regions of high Galactic latitude. Despite similarities exist in their oscillation modes, there are also discernible differences between $\delta$ Scuti stars and SX Phe stars. Compared with $\delta$ Scuti stars, SX Phe stars have on average lower luminosities and higher space velocities. Note that the SX Phe stars are usually discovered in globular clusters where blue stragglers are also prevalent \citep{cohen2012sx}. Whereas $\delta$ Scuti stars are frequently observed in the main sequence, indicating potential evolutionary distinctions, as SX Phe stars possibly originating from stellar mergers \citep{handler2009delta}. Furthermore, it's noteworthy that SX Phe stars exhibit periods ranging from 0.03 to 0.08 d (0.7-1.9 h, \citealt{kim2002differential}), statistically shorter than $\delta$ Scuti stars. As other pulsating stars, SX Phe stars follow $P-L$ relations. Given their presence in globular clusters, SX Phe stars can assist in determining the distances to globular clusters or dwarf galaxies \citep{mcnamara2011delta}. 

Figure \ref{fig:Distribution} shows the distribution of the 11,638 variables from TMTS in Galactic coordinates. Several variables in the classes of ``DSCT'' and ``HADS'' are found at high Galactic latitudes (i.e., |b|>30$\degree$). Combining with the criteria of $-$1.5 $<$ [Fe/H] $<$ $-$1.0 \citep{mcnamara2011delta}, we identified 8 candidates of SX Phe stars, which are summarized in Table \ref{tab:SX Phe}. To illustrate their distribution in the Galactic coordinates, we provide their Galactic longitude (l) and Galactic latitude (b), as well as their [Fe/H] obtained by the LAMOST. As the [Fe/H] parameter can be extracted for less than 20\% of our sample presented in this paper, this posts constraints on our capacity to identify additional candidates of SX Phe stars. A forthcoming work (Chen L., et al., in prep.) will discuss these SX Phe candidates in detail.

\section{Discussions}
\label{sec:discussions}

\begin{figure}
    \centering
    \includegraphics[width=0.5\textwidth]{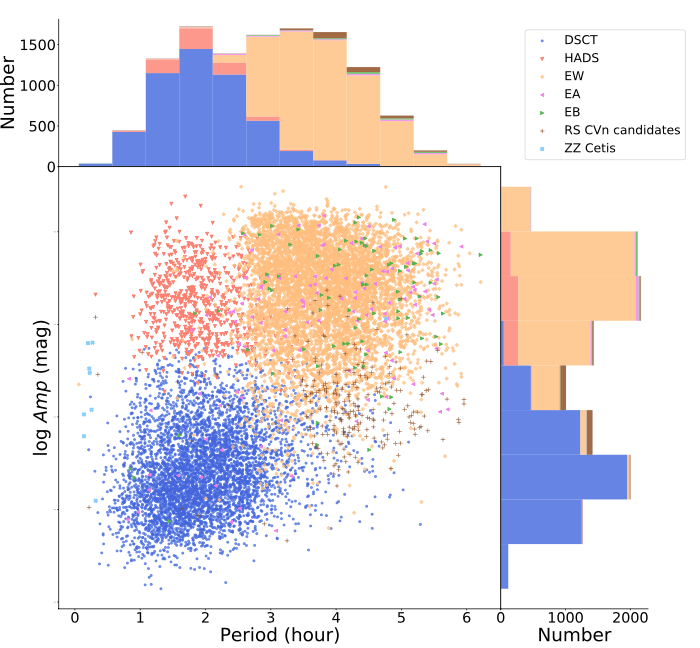}
    \caption{Relation between period and amplitude in logarithmic scale. DSCT, HADS and eclipsing binaries are well separated. A period histogram is at the top, and an amplitude histogram is on the right.}
    \label{fig:P-Amp}
\end{figure}

\begin{figure}
    \centering
    \includegraphics[width=0.49\textwidth]{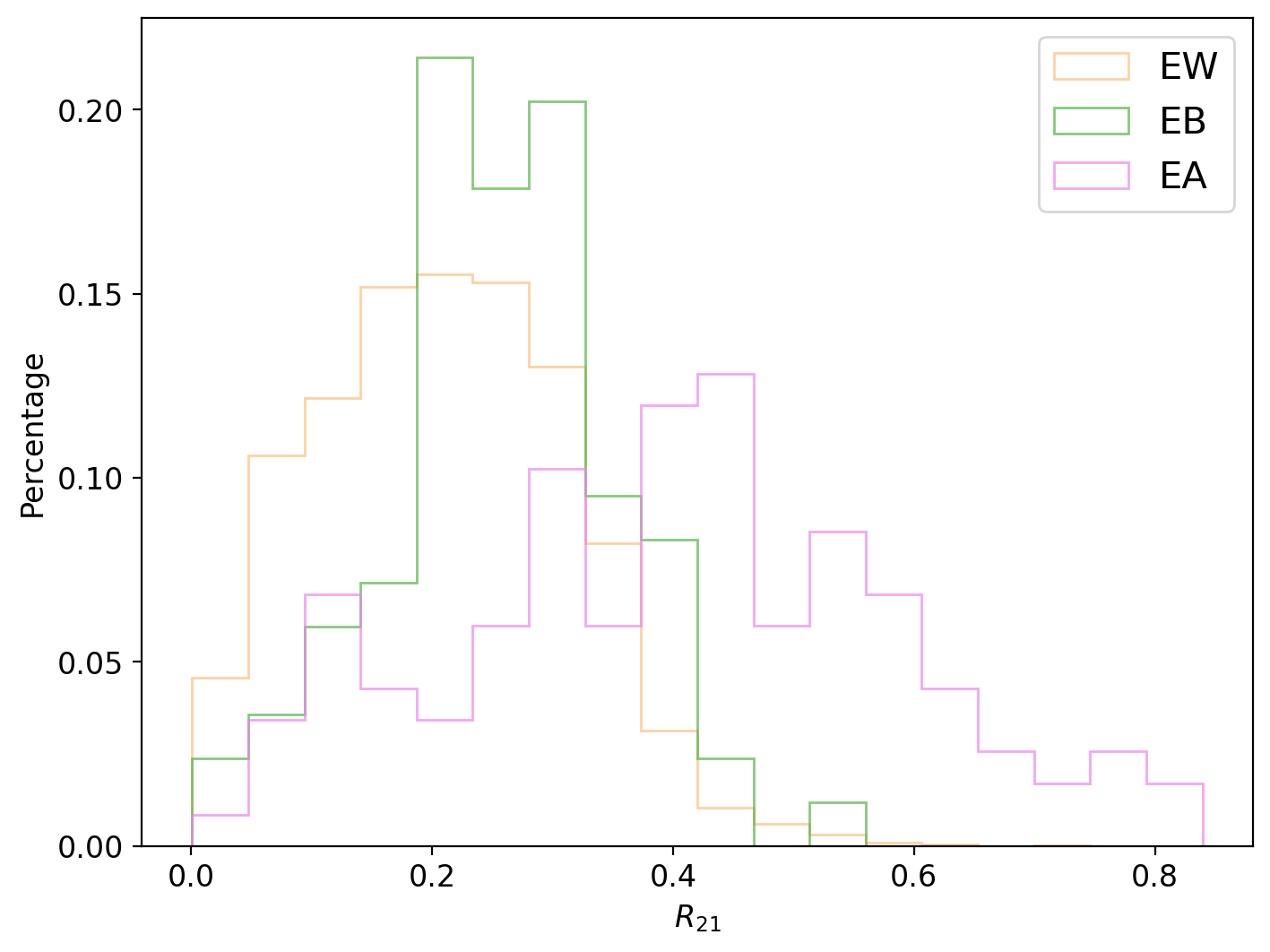}
    \caption{Histogram of $R_{21}$ of EWs, EBs and EAs. Larger $R_{21}$ is correlated with more asymmetric light curve. EAs statistically have largest $R_{21}$, followed by EBs and EWs.}
    \label{fig:R21}
\end{figure}

\begin{figure*}
    \centering
    \includegraphics[width=0.95\linewidth]{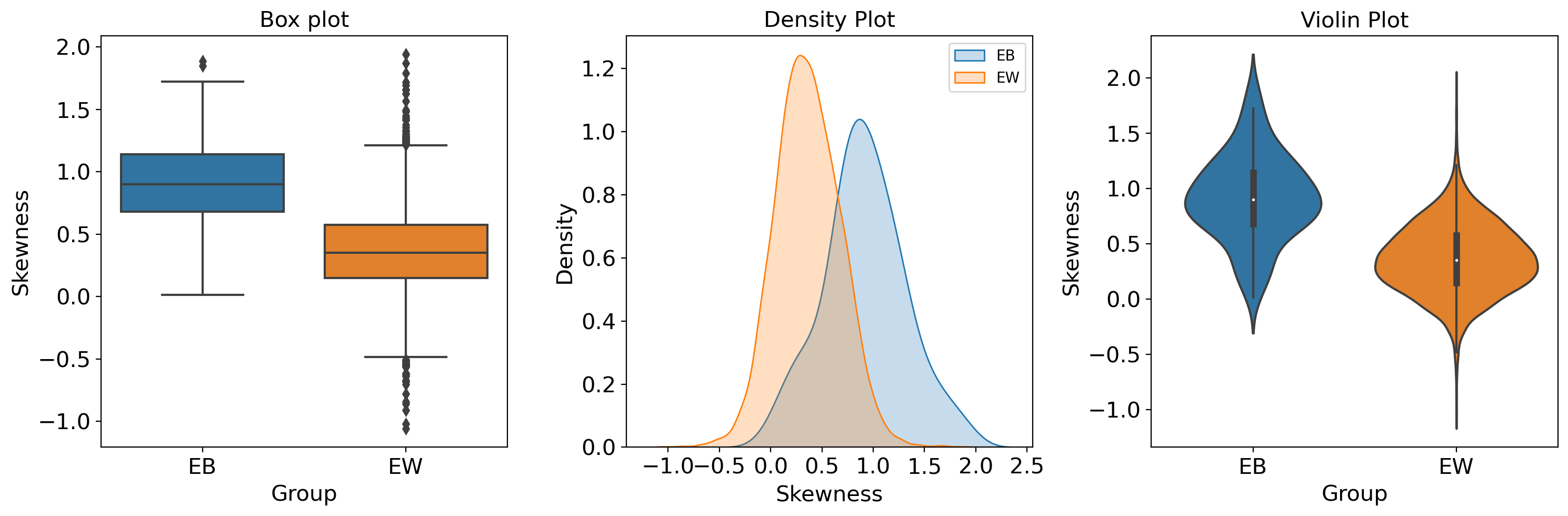} 
    \includegraphics[width=0.95\linewidth]{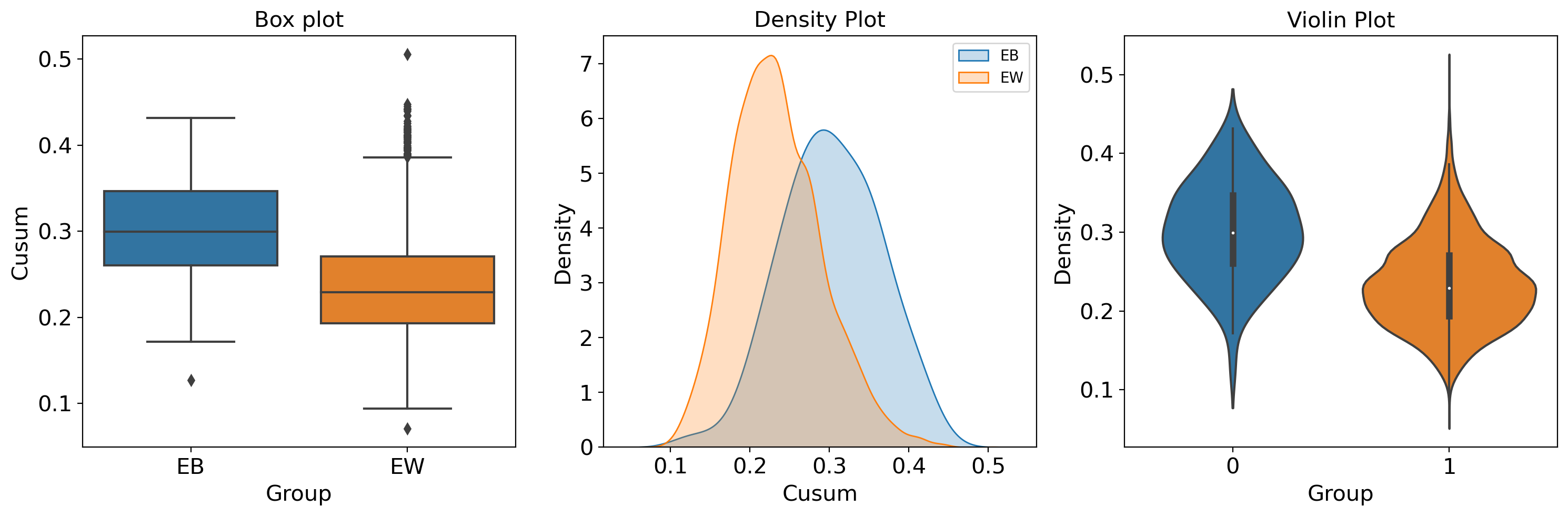}
    \caption{The box plot, density plot and violin plot depicting the distribution of skewness (upper) and cusum (lower) of the light curves of EBs and EWs. Box plot displays the minimum, first quartile, median, third quartile and maximum of a distribution, while density plot represents the probability density function of it. Violin Plot combines the information provided by the box plot and the density plot, using a kernel density estimation on each side.} 
    \label{fig:skewness&cusum} 
\end{figure*}

\begin{table} 
	\centering
	\caption{P-values of the K-S test and T-test on skewness, cusum and $R_{21}$ of EBs and EWs.}
	\label{tab:p value}
	\begin{tabular}{ccc} 
		\hline
            \hline
	       & K-S test & T-test\\
		\hline   
            Skewness & 2.068$\times10^{-32}$ & 2.235$\times10^{-54}$\\
            Cusum & 8.094$\times10^{-17}$ & 3.954$\times10^{-27}$\\
            $R_{21}$ & 7.827$\times10^{-5}$ & 4.050$\times10^{-5}$\\    
		\hline
	\end{tabular}
\end{table}

\begin{figure*}
    \centering
    \includegraphics[width=1\textwidth]{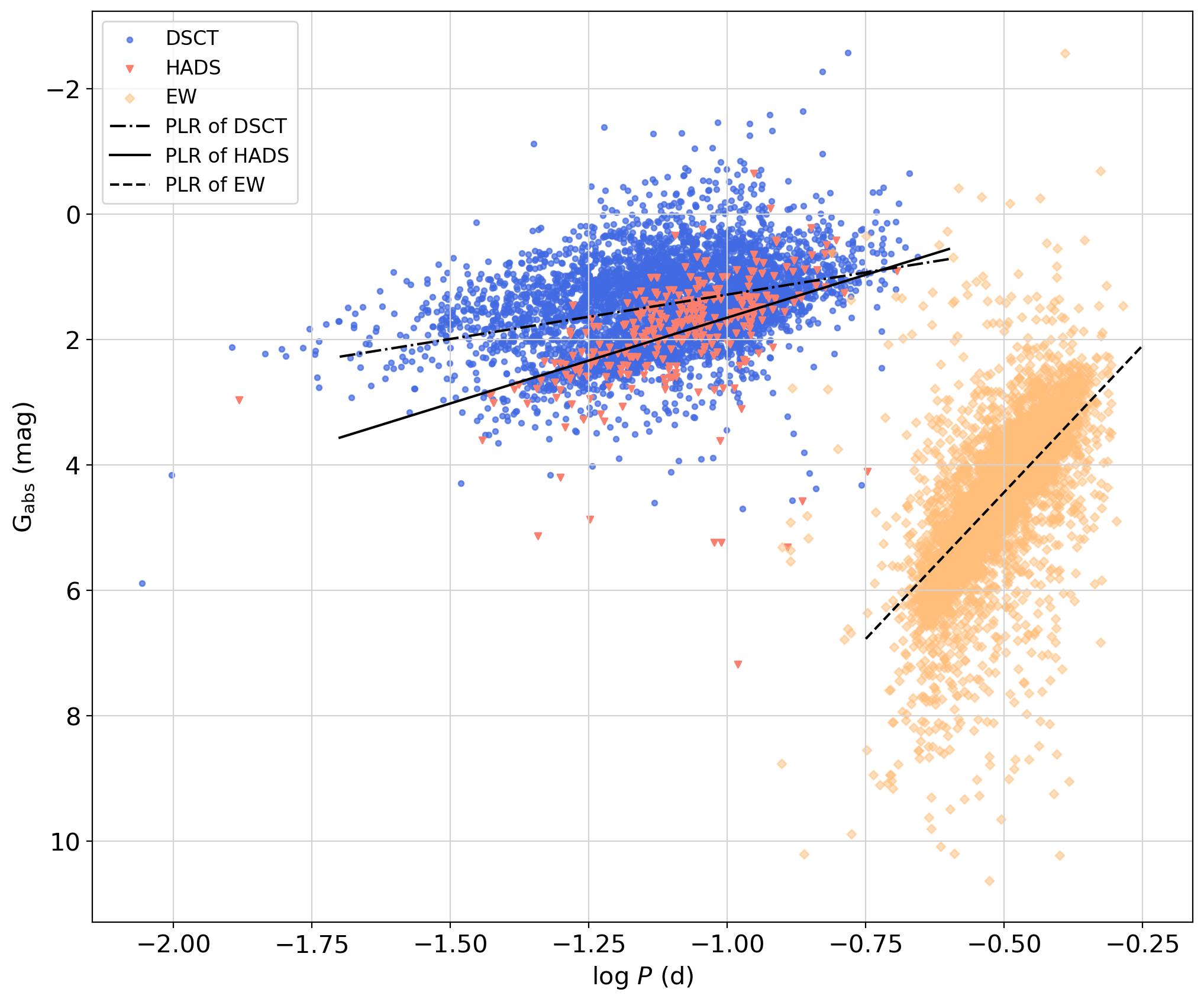}
    \caption{Period-Luminosity Relation of DSCT (dash-dotted line), HADS (solid line) and EW-type eclipsing binaries (dashed line). DSCT and HADS follow distinct $P-L$ relations, possibly due to different pulsation modes.}
    \label{fig:PLR}
\end{figure*}

\begin{figure*}
    \centering
    \includegraphics[width=1\textwidth]{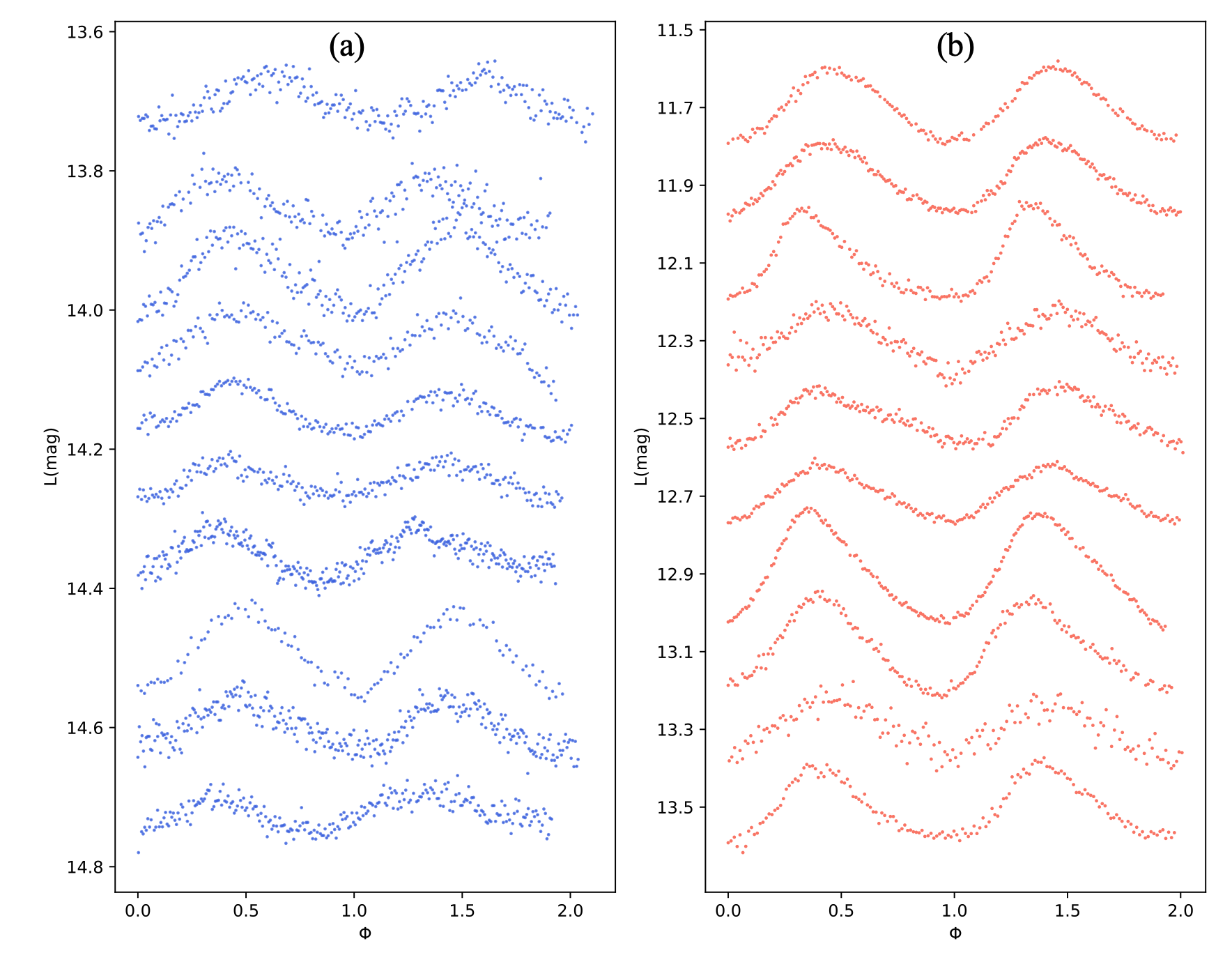}
    \caption{Example light curves of DSCT (a) and HADS (b). The mean magnitudes are fixed at 13.7+0.2i and 11.7+0.2i ($i$ = 0, 1, ..., 9) for DSCT and HADS, respectively.}
    \label{fig:dsct-hads}
\end{figure*}

We elaborate on the features used in the machine learning algorithms and the underlying physics, as well as discuss each type of variable stars in detail in this section, including their LC features and Period-Luminosity relations. 

\subsection{Comprehensive discussion of variable stars}
Figure \ref{fig:P-Amp} presents the relationship between period and amplitude of the variable stars while quantifying their distribution within each bin. This distribution allows for a rough differentiation between DSCT, HADS, eclipsing binaries and RS CVn candidates. DSCT and HADS typically have shorter periods, with HADS inherently possessing greater amplitudes. Conversely, eclipsing binaries and RS CVn candidates tend to have longer periods, while LCs of the former generally have larger light-variation amplitudes. As shown in Figure \ref{fig:Feature importance}, period is the most important feature in the classification process. The number of $\delta$ Scuti stars decrease with pulsating period \citep{rodriguez2001delta,qian2018lamost}, and most $\delta$ Scuti stars exhibit periods predominantly below 0.1 d. On the contrary, contact binaries have a short-period cutoff of 0.22 d (some studies have revised this cutoff to 0.15 d, \citealt{rucinski1992can,qian2020contact}), probably due to the fully convective limit \citep{rucinski1992can}, timescale of the angular momentum loss \citep{stepien2011evolution}, the low mass limit of the primary component \citep{jiang2012short}, etc. Although the cutoff-period refers to the orbital period (twice of the photometric period), most of the binary stars have photometric periods longer than the cutoff period. These trends suggest the presence of a natural period dichotomy between $\delta$ Scuti stars and binaries. It is noteworthy that the majority of $\delta$ Scuti stars in our catalog exhibit periods below 4 h, indicating the efficiency of TMTS in identifying short-period variables. 

In Figure \ref{fig:R21}, the Fourier parameter $R_{21}$ helps to separate EAs, EBs and EWs, especially EAs. $R_{21}$ measures the extent of deviation of a LC from a sinusoidal function, with higher values indicating more substantial deviations. Consequently, EAs exhibit the highest $R_{21}$ values, followed by EBs, while EWs tend to have the lowest values. Moreover, skewness and cusum are paramount contributors in distinguishing EBs from EWs. Skewness quantifies the degree of asymmetry within a distribution, while cusum measures its volatility. Elevated asymmetry and volatility correspond to increased values in both skewness and cusum, as observed in the light curves of EBs. The box plot, density plot and violin plot, depicting the distribution of skewness and cusum of the EBs and EWs in our catalog, are shown in Figure \ref{fig:skewness&cusum}. Furthermore, we conducted Kolmogorov-Smirnov (K-S) and T-tests on skewness, cusum and $R_{21}$ of EBs and EWs, with the correspinding p-values listed in Table \ref{tab:p value}. 
All p-values, including that of $R_{21}$, are noticeably below 0.0001, indicating a statistically significant distinction in the distribution of skewness, cusum and R21 between EBs and EWs at a significance level of 99.99\%.

The nature of eclipse is that the two stars periodically shade each other when orbiting around a common center of mass. As the process is symmetric, the LCs of eclipsing binaries tend to display a higher degree of symmetry. In contrast, driven by the $\kappa$ mechanism, the light curves of $\delta$ Scuti stars typically manifest rapid ascending phases followed by slower descents (saw-tooth-shaped), thereby introducing a higher level of asymmetry. As $\delta$ Scuti stars undergo contraction and expansion periodically, their opacity and luminosities change accordingly. Given that the contraction phase (corresponding to the decrease in the LC) is of a longer duration, the LCs of $\delta$ Scuti stars show a saw-tooth shape (\citealt{eddington1917pulsation,zhevakin1963physical}).

Given that various types of variable stars tend to occupy distinct regions within the $Gaia$ CMD, it becomes a powerful tool for classification. For instance, $\delta$ Scuti stars are typically found in the instability strip on or above the main sequence in the CMD. Although eclipsing binaries can appear anywhere, the observations conducted by TMTS (with upper-period-bound of 7.5 h) are primarily geared towards detecting $G$-, $K$-, or $M$-type stars (the CMD's lower right quadrant corresponds to stars with smaller radii, so that the distance between the two stars could be closer, which leads to shorter periods). This limitation naturally segregates eclipsing binaries from $\delta$ Scuti stars, which are typically comprised of $A$- to $F$-type stars.

\subsection{$\delta$ Scuti stars}

$\delta$ Scuti stars are $A0$- to $F5$-type stars pulsating in radial and non-radial acoustic modes. Their absolute magnitudes are typically below 3.5 mag, with temperature between 6900 K and 8900 K. $\delta$ Scuti stars locate in the intersection of the main sequence and the classical instability strip in the Hertzsprung–Russell diagram (as well as the $Gaia$ CMD). From a stellar evolutionary standpoint, most of them are main sequence stars in the hydrogen-burning stage, and a few may be zero-age main sequence (ZAMS) stars, red giants or blue stragglers moving off the main sequence. The $\kappa$ mechanism accounts for most oscillations seen in $\delta$ Scuti stars, where opacity increases with temperature in H and He ionisation zone. Their pulsation modes are complex, including low and intermediate order pressure modes (p-modes) and gravity modes (g-modes). Some stars are known to have multiple modes, which makes them promising objects for asteroseismological study. Among $\delta$ Scuti stars, HADS typically exhibit radial mode pulsation. With increased photometric precision, non-radial modes have also been detected in some cases. Compared with DSCT, HADS generally have lower rotational velocities (i.e., $v$ sin$i$ \(\leq\) 30 km$\mathrm{^{-1}}$), while $v$ sin$i$ can reach 200 km$\mathrm{^{-1}}$ for DSCT \citep{mcnamara1997luminosities,pigulski2006high}. 

Numerous models have provided theoretical calculations of the instability strip of $\delta$ Scuti stars \citep{dupret2004theoretical,xiong2016turbulent}. However, the theoretical edges does not match the observed pulsator fraction \citep{murphy2019gaia}. The new identification of over 4000 $\delta$ Scuti stars in our catalogue not only helps to better constrain the edges of the instability strip, but also offers valuable insights into the pulsator fraction inside the strip.

$P-L$ relations have been established for various pulsating stars \citep{leavitt1912periods}. Among them, the $P-L$ relations of Cepheids have been extensively studied. Like other pulsating stars, $\delta$ Scuti stars also follow $P-L$ relations \citep{breger1975period,mcnamara1997luminosities}, which offers an independent avenue for calibrating cosmic distances, complementary to the Cepheids. However, due to their intricate pulsation modes, lower luminosity and smaller amplitudes, the $P-L$ relations of $\delta$ Scuti stars has remained comparatively less well-defined \citep{ziaali2019period}. Figure \ref{fig:PLR} shows the P-L relations of DSCT, HADS and EW-type eclipsing binaries in our catalogues. We will discuss the $P-L$ relation of EWs later.
The best fitting $P-L$ relations of DSCT and HADS in our catalog are as follow:
\begin{gather}
    \mathrm{DSCT}: G\mathrm{_{abs}} = -1.42 \log(\frac{P}{\rm 1~d}) - 0.13\ \ ({\rm mag}) \\
    \mathrm{HADS}: G\mathrm{_{abs}} = -2.74 \log(\frac{P}{\rm 1~d}) - 1.08\ \ ({\rm mag}) 
\end{gather}

The increased dispersion observed in the $P-L$ relation of DSCT, as compared to HADS, can be ascribed to the fact that DSCT undergo multi-mode pulsations, encompassing both overtone mode and fundamental mode, whereas HADS predominantly exhibit fundamental mode pulsations. As expected, the overtone mode pulsation corresponds to a brighter $P-L$ relation compared to the fundamental mode (see also \citealt{lin2023minute}). Additionally, the overtone mode comprises a blend of various overtone modes, which may explain the larger dispersion and is consistent with \cite{jayasinghe2020asas}. In the short-period end of the period-luminosity diagram, the distribution of DSCT appears relatively sparse, forming what seems like two distinct lines, one aligns with the dash-dotted line ($P-L$ relation of DSCT), while the other corresponds to the solid line ($P-L$ relation of HADS). This may be attributed to the possibility that within the DSCT class, stars pulsating in the fundamental mode and overtone mode adhere to different $P-L$ relations. Conducting separate fittings of the $P-L$ relations for DSCT with distinct pulsation modes contributes to enhancing the accuracy of the $P-L$ relation. Further discussion and comparison of the $P-L$ relations will be undertaken in our future work (Chen L, et al., in prep.).

Figure \ref{fig:dsct-hads} shows some light curves of DSCT (a) and HADS (b). HADS commonly exhibit LCs characterized by a rapid rise and slow decline, while DSCT show a broader range of LC shapes, including more sinusoidal patterns. 

\subsection{Eclipsing Binaries}
\begin{figure*}
    \centering
    \includegraphics[width=1\textwidth]{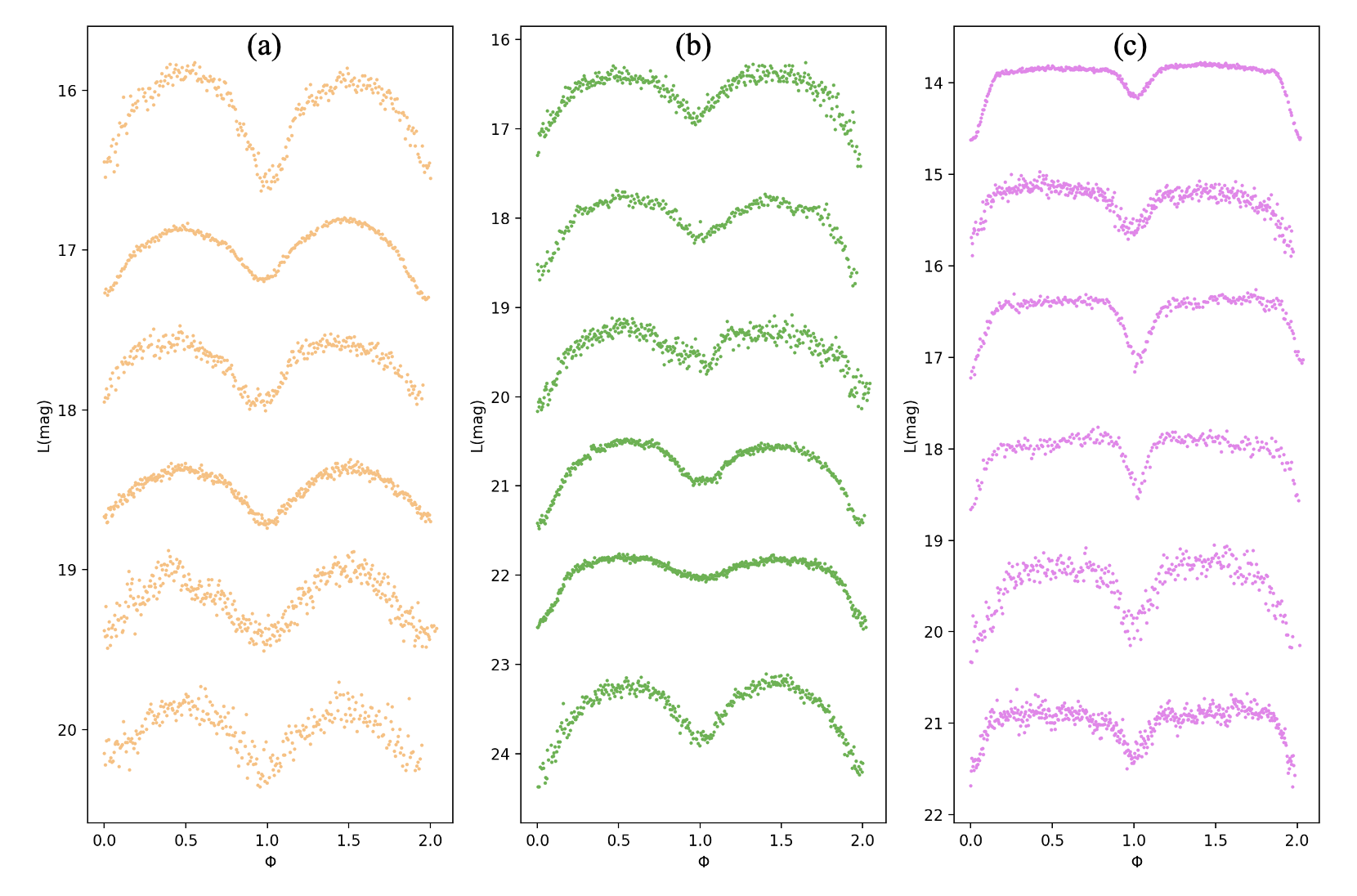}
    \caption{Example light curves of EW-type eclipsing binaries (a), EB-type eclipsing binaries (b) and EA-type eclipsing binaries (c). The mean magnitudes are fixed at 16.2+0.8i, 16.6+1.1i, and 14.2+1.5i ($i$ = 0, 1, ..., 5) for EWs, EBs and EAs, respectively.} 
    \label{fig:eclipsing}
\end{figure*}

Eclipsing binary exhibits eclipses due to the nearly edge-on orientation of its orbital plane with respect to the Earth, causing periodic changes in its brightness as one star passes in front of its companion. Unlike pulsating stars, they can appear anywhere in the CMD \citep{eyer2019gaia}. Eclipsing binaries offer unique opportunities for investigating material exchange and common envelope evolution due to the intricate interactions between the two components \citep{stassun2014empirical,pols1997further,nelson2023critical}. 

Eclipsing binaries can be categorized into three types based on their LC shapes, among which EWs accounts for the highest proportion, as illustrated in Figure \ref{fig:pie} and Table \ref{tab:classification}. Figure \ref{fig:eclipsing} displays typical light curves of the EWs (a), EB(b) and EAs (c). EAs exhibit clearly defined moments of beginning and end of the eclipse. Typically, there is minimum material exchange between the two spherical or slightly ellipsoidal stars, and they may be detached or semi-detached. In contrast, stars in the EB-type systems have active mass transfer between the envelopes, leading to continuous LC variations that make it impossible to determine the exact start and end moments of the eclipse. Additionally, EBs are characterized by different primary and secondary minimum depths, which distinguishes them from EW-type binaries. Finally, EWs are marked by nearly equal depths in their primary and secondary minima. Both components fill their Roche Lobes in EW-type systems. In addition, the densely sampled TMTS LCs would greatly contribute to the determination of minima epoch of eclipsing binaries in our catalogues. 

EW-type eclipsing binaries also follow a $P-L$ relation, especially in infrared bands, owing to the strong geometric constraints imposed by the common envelope \citep{rucinski2004contact,ren2021gaia}. In simple terms, orbital periods are correlated with radius of the system, and therefore with its luminosity. This $P-L$ relation renders EWs as reliable distance indicators. Compared to Cepheids and RR Lyrae stars, EWs trace older stellar population than Cepheids and are more common in open clusters and the solar neighborhood than RR Lyrae stars. This highlights their unique value as distance tracers \citep{chen2016contact}. However, owing to the limited sample size, the determination of $P-L$ relations for EWs is characterized by less precision, which impedes their application. \cite{chen2018optical} studied 27,318 EWs with accurate distance measurements, and they reached an distance accuracy of 8\%. Figure \ref{fig:PLR} shows the $P-L$ relation of EWs in our catalogue, the best fitting is 
\begin{equation}
    G\mathrm{_{abs}} = -9.36 \log(\frac{P}{\rm 1~d}) - 0.25\ \ ({\rm mag}) 
\end{equation}
Orbital period is used in this fitting. The $P-L$ relation of EWs may be influenced by factors such as color and metallicity \citep{rucinski2004contact}, and we only provide the simplest analysis in this paper. 

\subsection{RS CVn}
\begin{figure*}
    \centering
    \includegraphics[width=0.5\textwidth]{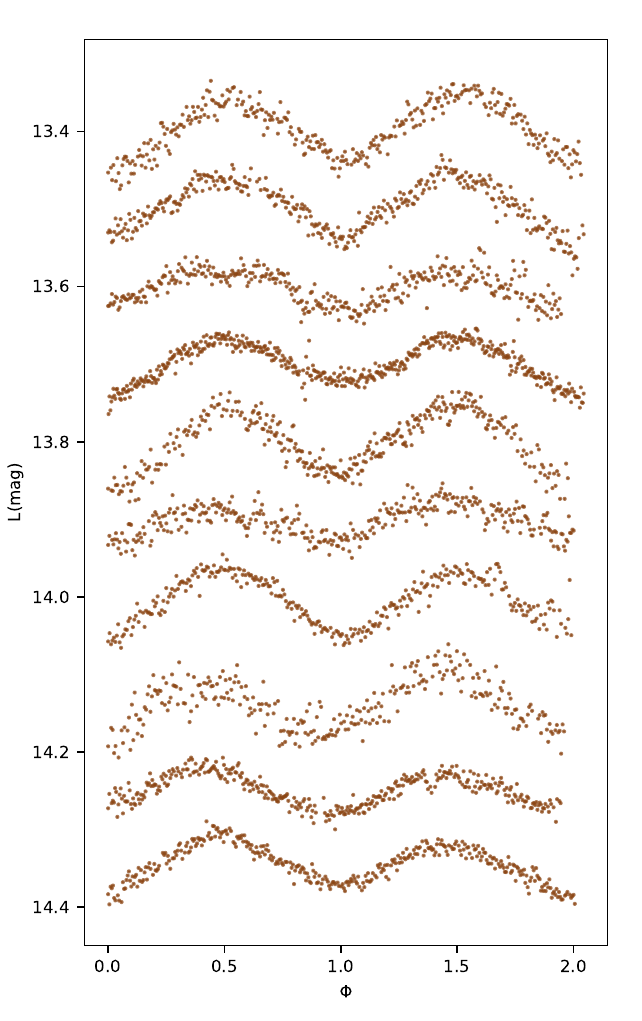}
    \caption{Example light curves of RS CVn candidates. The mean magnitudes are fixed at 13.4+0.1i ($i$ = 0, 1, ..., 9).}
    \label{fig:rs}
\end{figure*}

RS CVn variables are an active subclass of binary systems characterized by strong Ca II H \& K emission lines emerged from at least one of the system's components \citep{hall1976rs}. Consequently, the periodic photometric variations in these systems arise not only from pulsations or eclipses, but also from their rotation. The primary star of RS CVn is a F to K-type subgiant or giant star, with a G to M-type dwarf or subgiant companion \citep{martinez2022activity,fekel1986survey}. The orbital periods of RS CVn variables exhibit variability over time, and studying these changes would shed light on the mass loss process, magnetic field interactions and the dynamic evolution of binary systems \citep{hall1980period}. It's crucial to mention that a comprehensive understanding of RS CVn variables requires detailed spectral properties. Therefore, in this paper, we only present a catalogue of RS CVn candidates.

The LCs of RS CVn variables exhibit a discernible semiperiodic pattern, which coexists alongside the eclipse features. This phenomenon is attributed to the presence of ``starspots''. Arising from strong magnetic fields and stifled convection, these starspots are cooler regions on the photosphere in the outer layer of the star \citep{roettenbacher2015detecting}. Figure \ref{fig:rs} shows the LCs of some RS CVn candidates in our catalogue.

\section{Summary}
\label{sec:summary}
With a cadence of \textasciitilde 1 minute, TMTS has shown great potential in the detection of short-period variables. In this paper, we systematically classified 11,638 variable stars into 6 main categories using the state-of-the-art XGBoost and Random Forest algorithms, achieving remarkable accuracy levels of 98.83\% and 98.73\%, respectively. We constructed a well-chosen training and test dataset of 4506 light curves of the variables. Period is the most important feature in the classification, followed by features that characterize the shape of the LCs as well as $Gaia$ $G$$\mathrm{_{abs}}$ and $(B\mathrm{_p}-R\mathrm{_p})_0$, which is in accordance with our predictions. Once the effectiveness of the classifiers was confirmed, we utilized them to assign labels to the remaining unlabeled light curves. Notably, in both classifiers, the macro-average and weighted-average F$_1$ scores are higher than 98\%, with the F$_1$ scores for DSCT, EA, EB and HADS exceeds 99\%, demonstrating the effectiveness of our feature selection and model training methodologies. To ensure the quality of the catalogue, we conducted a manual inspection to all included light curves. The period-amplitude diagram and the period-$R_{21}$ diagram prove the validity of the classification.

We identified 5504 $\delta$ Scuti stars (including 4876 typical $\delta$ Scuti stars and 628 high-amplitude $\delta$ Scuti stars), 5899 eclipsing binaries (including 117 EAs, 84 EBs and 5698 EWs) and 226 candidates of RS Canum Venaticorum systems. Additionally, with the help of color-magnitude diagram established from the Gaia database, we discovered 9 ZZ Ceti stars. Combining Galactic longitudes with metal abundances, we discovered 8 metal-poor $\delta$ Scuti stars (SX Phe stars). Among the 4876 DSCT in our catalog, 4297 were newly discovered, and most of them have periods shorter than 4 hours, showing the ability of TMTS to search for short-period variables. The newly-identified $\delta$ Scuti stars could help to better refine the red and blue edges of the instability strip. We study the period-luminosity relations of DSCT, HADS and EW-type eclipsing binaries. The subclasses of DSCT and HADS follow distinct $P-L$ relations, and different $P-L$ relations seem to also exist within the class of DSCT, potentially attributable to the difference in fundamental mode and overtone mode pulsations. An in-depth analysis of $\delta$ Scuti stars with varying pulsation modes would contribute to a more precise determination of the $P-L$ relation. We will further discuss DSCT, HADS and their $P-L$ relations in the forthcoming paper.

\section*{Acknowledgements}
This work is supported by the National Science Foundation of China (NSFC grants 12033003, 12288102, and 11633002), the Ma Huateng Foundation, the Scholar Program of Beijing Academy of Science and Technology (DZ:BS202002), and New Cornerstone Science Foundation through the XPLORER PRIZE. J.L. is supported by the Cyrus Chun Ying Tang Foundations. Y.-Z. Cai is supported by the National Natural Science Foundation of China (NSFC, Grant No. 12303054) and the International Centre of Supernovae, Yunnan Key Laboratory (No. 202302AN360001).

This work is supported by the National Natural Science Foundation of China (NSFC) grant 12373031, the Joint Research Fund in Astronomy (U2031203) under cooperative agreement between the National Natural Science Foundation of China (NSFC) and Chinese Academy of Sciences (CAS), and the NSFC grants 12090040, 12090042. This work is also supported by the the CSST project.

We acknowledge the support of the staffs from Xinglong Observatory of NAOC during the installation, commissioning, and operation of the TMTS system. 

Guoshoujing Telescope (the Large Sky Area Multi-Object Fiber Spectroscopic Telescope LAMOST) is a National Major Scientific Project built by the Chinese Academy of Sciences. Funding for the project has been provided by the National Development and Reform Commission. LAMOST is operated and managed by the National Astronomical Observatories, Chinese Academy of Sciences.

This research has used the services of www.Astroserver.org. This work has made use of data from the European Space Agency (ESA) mission Gaia \url{(https://www.cosmos.esa.int/ gaia)}, processed by the Gaia Data Processing and Analysis Consortium (DPAC, \url{https://www.cosmos.esa.int/web/gaia/dpac/consortium}). Funding for the DPAC has been provided by national institutions, in particular the institutions participating in the Gaia Multilateral Agreement.

This work has made use of data from the the Optical Gravitational Lensing Experiment (OGLE) and the All Sky Automated Survey for SuperNovae (ASAS-SN). The OGLE project has received funding from the National Science Centre, Poland (grant number MAESTRO 2014/14/A/ST9/00121 to A.U.). The ASAS-SN is supported at OSU by the Gordon and Betty Moore Foundation through grant GBMF5490 to the Ohio State University and NSF grant AST-1908570.

This research has made use of the International Variable Star Index (VSX, \citealt{watson2006international}) database, operated at AAVSO, Cambridge, Massachusetts, USA. Some of the results in this paper have been derived using the HEALPix \citep{gorski2005healpix} package.

\section*{Data Availability}
The catalogues are all available from this paper and CDS.




\bsp	
\label{lastpage}
\end{document}